\documentclass[12pt]{article}

\usepackage{amsmath}
\usepackage{amssymb} 
\usepackage{makeidx}
\usepackage{epsf}
\usepackage{graphicx}
\begin{document}
 \newcommand{\rot}{\mathop{\rm rot}\nolimits}
 \newcommand{\grad}{\mathop{\rm grad}\nolimits}
 \newcommand{\diver}{\mathop{\rm div}\nolimits}
 \newcommand{\km}{\mathop{\rm km}\nolimits}
 \newcommand{\s}{\mathop{\rm s}\nolimits}
 \newcommand{\keV}{\mathop{\rm keV}\nolimits}
 \newcommand{\eV}{\mathop{\rm eV}\nolimits}
 \newcommand{\Ma}{\mathop{\rm Ma}\nolimits}
 \newcommand{\const}{\mathop{\rm const}\nolimits}
 \newcommand{\Erf}{\mathop{\rm Erf}\nolimits}
\newcommand{\Ree}{\mathop{\rm Re}\nolimits}
 \normalsize

\section*{TURBULENCE IN CLUSTERS OF GALAXIES AND X-RAY LINE PROFILES}

\centerline{N.A. Inogamov$^{1,2}$ and R.A. Sunyaev$^{2,3}$}

\small

$^1$ Landau Institute for Theoretical Physics,
Russian Academy of Sciences,
Chernogolovka, Russia

\noindent $^2$ Max-Planck-Institut f\"ur Astrophysik, Garching, Germany

\noindent $^3$ Space Research Institute,
Russian Academy of Sciences,
Moscow

\normalsize

\vspace{.3cm}

\noindent Full text of the paper will be published in December, 2003 issue
of Astronomy Letters {\bf 29,} 791.

\vspace{.3cm}

{\bf Abstract}
Large-scale bulk motions and hydrodynamic turbulence
in the intergalactic gas
inside clusters of galaxies
significantly broaden X-ray emission lines.
For lines of heavy ions
(primarily helium-like and hydrogen-like iron ions),
the hydrodynamic broadening is significantly larger
than the thermal broadening.
Since cluster of galaxies have a negligible optical depth
for resonant scattering
in forbidden and intercombination lines of these ions,
these lines are not additionally broadened.
At the same time, they are very intense,
which allows deviations of the spectrum
from the Gaussian spectrum in the line wings
to be investigated.
The line shape becomes an important indicator
of bulk hydrodynamic processes
because the cryogenic detectors
of new generation of X-ray observatories
will have a high energy resolution
(from 5 eV for ASTRO-E2 to 1-2 eV for Constellation-X and XEUS).
We use the spectral representation of a Kolmogorov cascade
in the inertial range
to calculate the characteristic shapes of X-ray lines.
Significant deviations in the line profiles from the Gaussian profile
(shape asymmetry, additional peaks, sharp breaks in the exponential tails)
are expected for large-scale turbulence.
The kinematic SZ effect
and the X-ray line profile carry different information
about the hydrodynamic velocity distribution in clusters of galaxies
and complement each other,
allowing the redshift,
the peculiar velocity of the cluster,
and the bulk velocity dispersion to be measured and separated.

\vspace{.1cm}

\noindent {\bf Key words:} {\it turbulence, clusters of galaxies, intergalactic gas,
X-ray line spectroscopy.}

\vspace{.2cm}

\section{The General Picture}

The hot intergalactic gas in clusters of galaxies
forms an extended atmosphere in the gravitational potential well
produced mainly by the weakly interacting dark matter,
whose mass exceeds the intergalactic gas mass
by a factor of approximately six.
The cluster gas temperature reaches 2-10 KeV.
The speed of sound in this gas is 1000-1500 km s$^{-1}.$
Galaxies move through the cluster gas
with subsonic, sonic and supersonic velocities.
Energetically, the presence of galaxies is of little importance,
because the total mass of the galaxies
is appreciably smaller than the total cluster mass.
The intergalactic gas consists of completely ionized hydrogen
together with $\sim$ 25 \% (by mass) of helium.
Heavy elements, including iron,
are represented at a 30 - 40\% level of their solar abundance.

Due to the growth of large-scale cosmological density perturbations,
clusters of galaxies sometimes merge together
and capture surrounding galaxies and groups of galaxies.
During such merger,
the colliding components move with supersonic or transonic velocities.
Each merger is accompanied by shock wave formation
and turbulence generated on smaller scales.
The characteristic time between mergers is long.
The effects of no more than one to three large mergers
appear to be observed simultaneously;
a large merger is a merging of clusters with comparable masses.
The capture of low-mass companions or individual galaxies
gives a smaller contribution to the generation of turbulence.

It follows from $N$-body simulations
with cold dark matter and gas
that the turbulent pressure can account for
up to 15\% of the thermal pressure
of the intergalactic gas in relaxed clusters of galaxies.
This value implies very high turbulent pulsation velocities
(up to 40\% of the speed of sound).
Remarkably, the calculations by different groups
yield similar results
(the methods, codes, and results
were carefully compared by Frenk et al. 1999;
see also Norman and Bryan 1999).

Activity of accreting supermassive black holes
in the dominant galaxy
inside the cluster
can trigger additional small-scale turbulence
due to existence of collimated high-velocity mass outflows
and interaction of intergalactic gas
with relativistic jet (Churazov et al. 2002a, 2002b).

\subsection{X-ray Line Profiles}

In 2005, ISAS and NASA are planning to launch the ASTRO-E2 satellite\footnote{
http://www.isas.ac.jp/e/enterp/missions/astro-e2/}.
This satellite
will be equipped with cryogenically cooled X-ray bolometers.
They will be placed
in the focal planes of grazing-incidence telescopes.
The expected energy resolution
will be $\sim$ 5 eV (FWHM)
in the photon energy range from 0.5 to 10 keV,
which includes the line of the helium-like iron ion
(the iron ion with two electrons)
at energy 6.7 keV.
Since below we compare the line profiles
with Gaussian fits,
it is important to note
(Porter and Mitsuda 2003)
that this resolution corresponds to a dispersion of 2.7 eV.
The prospective Constellation-X\footnote{
http://constellation.gsfc.nasa.gov/docs/main.html}
and XEUS\footnote{
http://astro.estec.esa.nl/SA-general/Projects/XEUS/}
missions
will have X-ray detectors
with an energy resolution as high as 1-2 eV.
Therefore, it will become possible
to detect line shifts and broadenings
in rich cluster of galaxies
that correspond to turbulent velocities
$\sim$ 3 $\times$ 10$^{-4}$ of the speed of light,
i.e. up to 100 km s$^{-1}$
or less than 10\% of the speed of sound.

During turbulent pulsations,
iron ions move in a mixture with hydrogen and helium nuclei.
Accordingly, all ions have the same hydrodynamic velocities.
The thermal velocities of ions of different types
with the same temperature
greatly differ
because of the large difference between the nuclear masses.
Therefore, the Doppler turbulent broadening of iron lines
$\Delta E/E \sim v_{\rm turb}/c$
can significantly exceed their thermal broadening.
The thermal velocity of iron ions
is $\sim \sqrt{m_p/m_{Fe}} = 1/\sqrt{56}$ = 13\% of the speed of sound,
where $m_p$ and $m_{Fe}$
are the masses of the proton and the iron atomic nuclei,
respectively.
The amplitude of the turbulent velocity pulsations
produced by a cluster merger
can exceed the thermal velocities of iron ions several fold.
Thus, iron ions become effective tracers of turbulent velocity fields
in the cluster interiors.

Let us introduce the parameter
$$
\Ma_i=\frac{u_{\rm rms}}{\sqrt{ kT/m_i} } =
 \sqrt{2}\,
\frac{\Delta\nu_{\rm DH}}{\Delta\nu_{\rm DT}}
 = 4.2\, \frac{u_{\rm rms}}{300\, [\,\rm {km \; s^{-1}}\, ]
}\,
\sqrt{\frac{3\,[\rm keV]}{kT}}\,
\sqrt{\frac{m_i}{56 \, m_p}}
\eqno (1.1)
$$
that characterizes the ratio of the turbulent
($v_{\rm turb}\sim u_{\rm rms}$)
and thermal velocity scales for an ion of mass $m_i$.
It resembles the Mach number:
$Ma = u_{\rm rms}/c_s$
(the ionic Mach number),
where $c_s$ is the speed of sound.
Parameter (1.1)
defines the ratio of the hydrodynamic ($\Delta\nu_{\rm DH}$)
and thermal ($\Delta\nu_{\rm DT}$) Doppler broadenings:
$$
\Delta\nu_{\rm DH} = \nu_0 \,\frac{u_{\rm rms}}{c}, \;\;\;\;\;
\Delta\nu_{\rm DT} = \nu_0 \,\frac{\sqrt{2kT/m_i}}{c}.
\eqno (1.2) 
$$
In (1.1) and (1.2) $u_{\rm rms}$
is the line-of sight or radial hydrodynamic velocity dispersion
and $\nu_0$ is the frequency at the center of the line profile.

The most intense lines
observed in the spectral range between 2 and 10 keV
are those of helium and hydrogen-like iron ions.
The equivalent width of the most intense iron lines
reaches 100-500 eV.
At the thermal Doppler width of $\sim$ 3 eV,
they must rise by tens and hundreds of times
above the smooth continuum
associated with hydrogen-helium plasma bremsstrahlung.
XMM observations
of the central part of the Perseus cluster of galaxies
gave 10$^5$ photons in 50 ks in the complex of iron K-lines
near 6.7 keV
(Churazov et al. 2003).
The Constellation-X and XEUS satellites
will have an effective area
that is tens of times larger
than that of XMM,
which will make it possible
to study in detail the weak wings of X-ray lines.

The spectral surface brightness of a cluster in a line
is given by the formula 
\[
I(\Delta E,y,z)=\int dx\,j[E_0, T_e(x,y,z) ]\,
 n_e(x,y,z)\, \frac{d n_i (x,y,z,\Delta E)}{d(\Delta E)},
\]
where $I$ is measured in
photons cm$^{-2}$ s$^{-1}$ eV$^{-1}$ (square angular minute)$^{-1},$
the emissivity coefficient $j$
characterizes the rate of production of excited ions
and emission of photons with energy $E_0$ by these ions,
$n_e$ is the electron density,
and $dn_i/d(\Delta E)$ is the number of ions of a given type
in the range of velocities
that shift the line photons to the energy range
from $\Delta E {\rm to} \Delta E + d(\Delta E$)
(in the frame of reference associated with the ion,
the photon energy is $E_0,$
without Doppler and intrinsic broadenings).
The energy $ \Delta E = E - E_0$
is measured from the line center
and is proportional (see below) to the line-of-sight ion velocity.
The line-of-sight velocity of ions
can be determined
by taking into account their thermal and hydrodynamic velocities.
We assume that the thermal velocity distribution of ions
is Maxwellian
with an ion temperature equal to the electron temperature.
The hydrodynamic velocity distribution
can be calculated in terms of the turbulence model presented below.
In the above formula,
the natural line width is disregarded
(see formulas (12.2) - (12.5) in the full paper).
Emissivity coefficient $j$
depends on the electron temperature $T_e.$
In an isothermal plasma
at a constant (in volume) elemental abundance,
$n_e n_i \propto n_e^2.$

In this paper, we consider qualitative effects,
primarily
the line broadening,
its spectral shift due to bulk and turbulent motions
and
the degree of deviation of the emission line profile
from the Gaussian profile.
Therefore, for illustration,
we investigate the simplest case of an isothermal cloud
with constant density and elemental abundance
over its volume.
In this case, the above formula for the brightness $I$ takes the form
\[
I(\Delta E,y,z)=j(E_0, T_e)\, n_e \int dx\,\frac{d n_i (\Delta
E)}{d(\Delta E)}=
\alpha\,j \, n_e^2\, \int dx\, \psi (x,y,z,\Delta E),
\]
where $\alpha$ is the elemental abundance,
and the function $\psi$
specifies the local line-of-sight velocity distribution.
We ignore the nonlinear density variations due to turbulent pulsations.

An efficient method for solving the problem of the iron line profile
involves direct numerical simulation.
The first attempt of this kind was made by
Sunyaev, Norman, and, Bryan (2003).
Below, we attempt to construct a relatively simple turbulence model
(see formula (4.1) below)
to understand
in which cases {\it the shape of a turbulently broadened line
can differ from the Gaussian shape
expected from the central limit theorem of the probability theory.}
The model is based on the assumption about a Kolmogorov cascade
in the inertial range
limited by the mixer size $L_{\rm mix}$
and the viscous scale $L_\nu$.
The number of mixers in the large-scale case ($L_{\rm mix} \sim L$)
is limited.
As a result,
the line spectrum significantly deviates from the Gaussian spectrum\footnote{
The influence of the number of mixers
on the deviation from Gaussian distribution
is discussed in Section 7, see also the full paper.
Large-scale motion shifts the line center, Fig. 2 and Table 1 below.
The presence of a Kolmogorov tail in small scales
does not imply
that the profile will be Gaussian (see Table 2).
}.

\begin{figure}
\mbox{}\hfill\epsfxsize 0.7\hsize\epsfbox{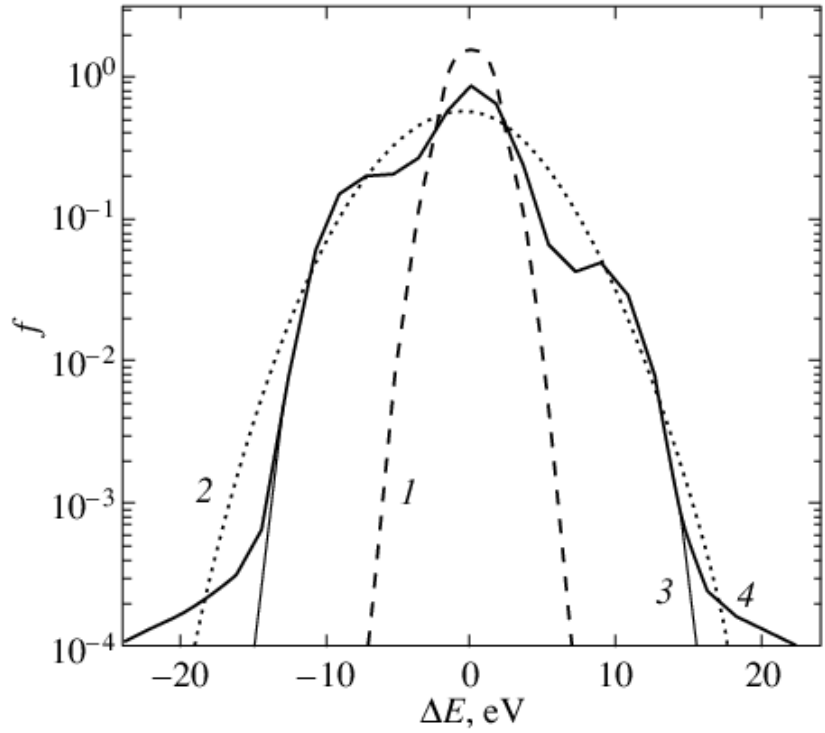}\hfill\mbox{}
\caption{
Fe XXV $w$-line spectrum (4) at $kT$ = 3 keV;
1 -- the thermally broadened line profile
(without hydrodynamic broadening),
2 -- the Gaussian profile with a total
(hydrodynamic plus thermal)
dispersion $u_{\rm HT}$ = 190 km s$^{-1}$,
and 3 -- the line profile without Lorentz broadening.
\label{fig:fig1}
}
\end{figure}

Fig. \ref{fig:fig1}
shows the characteristic profile
of the helium-like iron
(Fe XXV)
permitted $w-$line (curve 4)
obtained in the adopted model.
As we see (cf. curves 2 and 4),
there are qualitative deviations
from the commonly assumed Gaussian line profile
\[
\frac{ u_{\rm rms}
\exp\left[ - (1/2) \, u^2 / ( u_{rms}^2 + kT/m_i ) \right]
}{
\sqrt{2\pi}\,
\sqrt{ u_{rms}^2 + kT/m_i }
}
=
\frac{
(2\pi)^{-1/2}
}{\sqrt{ 1 + \Ma_i^{-2} }}\,
\exp \left( - \frac{ (1/2) \, \hat u^2 }{ 1 + \Ma_i^{-2} } \right)
\]
with a dispersion
the square of which is equal to the sum
of the squares of the thermal and hydrodynamic dispersions
(curve 2).
In what follows
$$
\hat{u} = u/u_{\rm rms},\;\;\;
\Delta E[\rm eV] = - 6.7\,[\rm eV] \,
(u_{\rm rms}/300[{\rm km \, s^{-1}} ]) \, \hat{u},
\eqno (1.3)
$$
$\hat{u}$ is a dimensionless velocity (in units of dispersion),
and $\Delta E$ is the deviation from the profile center.
Formula (1.3)
is used to recalculate $\hat{u}$ to $\Delta E$.
Parabola 1 in Fig \ref{fig:fig1} corresponds to the Maxwellian distribution of ions
$
f_T (\hat{u}) =
(Ma_i/\sqrt{2\pi})
\exp ( - \Ma_i^2\,\hat{u}^2/2$)
(purely thermal broadening).
All distributions where normalized to unity
$\int^{\infty}_{-\infty} f(\hat{u}) d \hat{u}$ = 1.
Curves 3,
$f_{\rm HT }(\hat{u}$),
and 4,
$f_{\rm HTL}(\hat{u}$),
represent the turbulent line profiles.
The distribution $f_{\rm HT} (\hat{u}$)
includes hydrodynamic (H) and thermal (T) broadenings,
while the distribution $f_{\rm HTL}(\hat{u}$)
includes hydrodynamic, thermal, and Lorentz (L) broadenings.
The intrinsic or Lorentz broadening
was calculated for the Fe XXV $w-$line\footnote{
See calculations in Sections 12 and 13 in the full paper.
}.

\begin{figure}
\mbox{}\hfill\epsfxsize 0.99\hsize\epsfbox{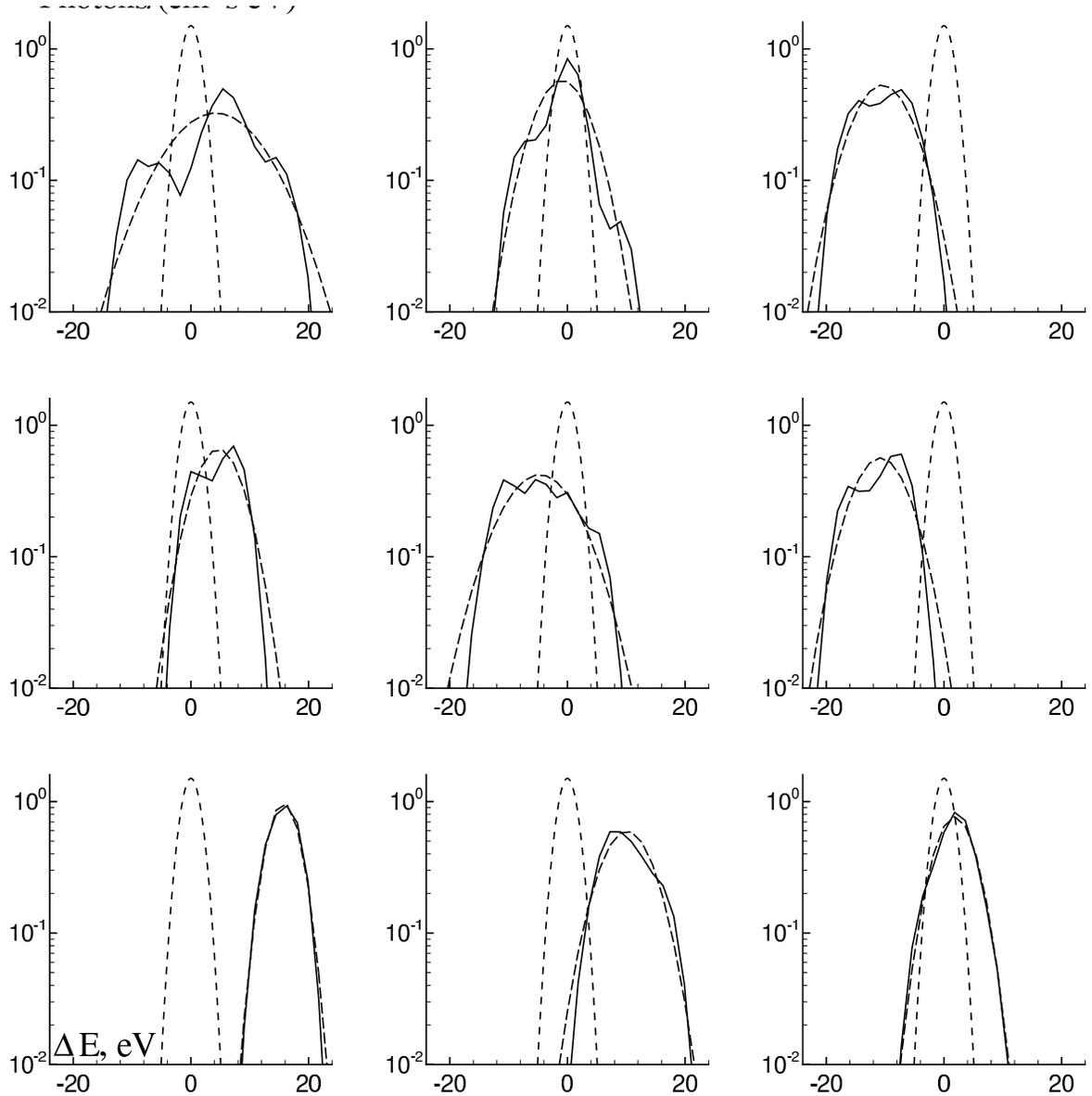}\hfill\mbox{}
\caption{
Helium-like iron line profiles
at nine points in the image of a cluster
with $kT$ = 3 KeV.
The solid curves
represent line profiles
shifted and broadened due to hydrodynamic and thermal motions
and the dashed curves
(narrow parabolas)
specify the thermal line broadening.
The hydrodynamic and thermal velocity dispersions
and the velocity
corresponding to the shift of the line profile centroid
are given in Table 1 for each of the nine lines.
The long-dashed curves (wide parabolas)
represent the Gaussian fits to the computed spectra.
\label{fig:fig2}
}
\end{figure}

Fig. \ref{fig:fig2} shows,
first,
the computed spectra
(fluctuating solid curves),
second,
the thermal Gaussians of the helium-like iron $w$-line
(dashed curves, narrow parabolas),
and, third,
the Gaussian fits to the computed spectra
(long0dashed wide parabolas).
These spectra were constructed for nine points
of the cluster projection onto the plane of the sky.
The Gaussian fits are specified by two parameters
(shift $u_1$ and broadening $u_{\rm HT})$
that can be determined from the computed spectrum
(see Table 1).
These nine columns
correspond to the center, vertices, and middles of the sides
of the square
in the plane of the sky described in Sections 1.2 and 1.3,
see also Section 11 of the full paper.
Note that curve 3 in Fig. \ref{fig:fig1}
corresponds to the line profile
shown in the middle panel in the upper row of Fig. \ref{fig:fig2}.

\begin{figure}
\hbox{
\includegraphics[width=0.5\columnwidth]{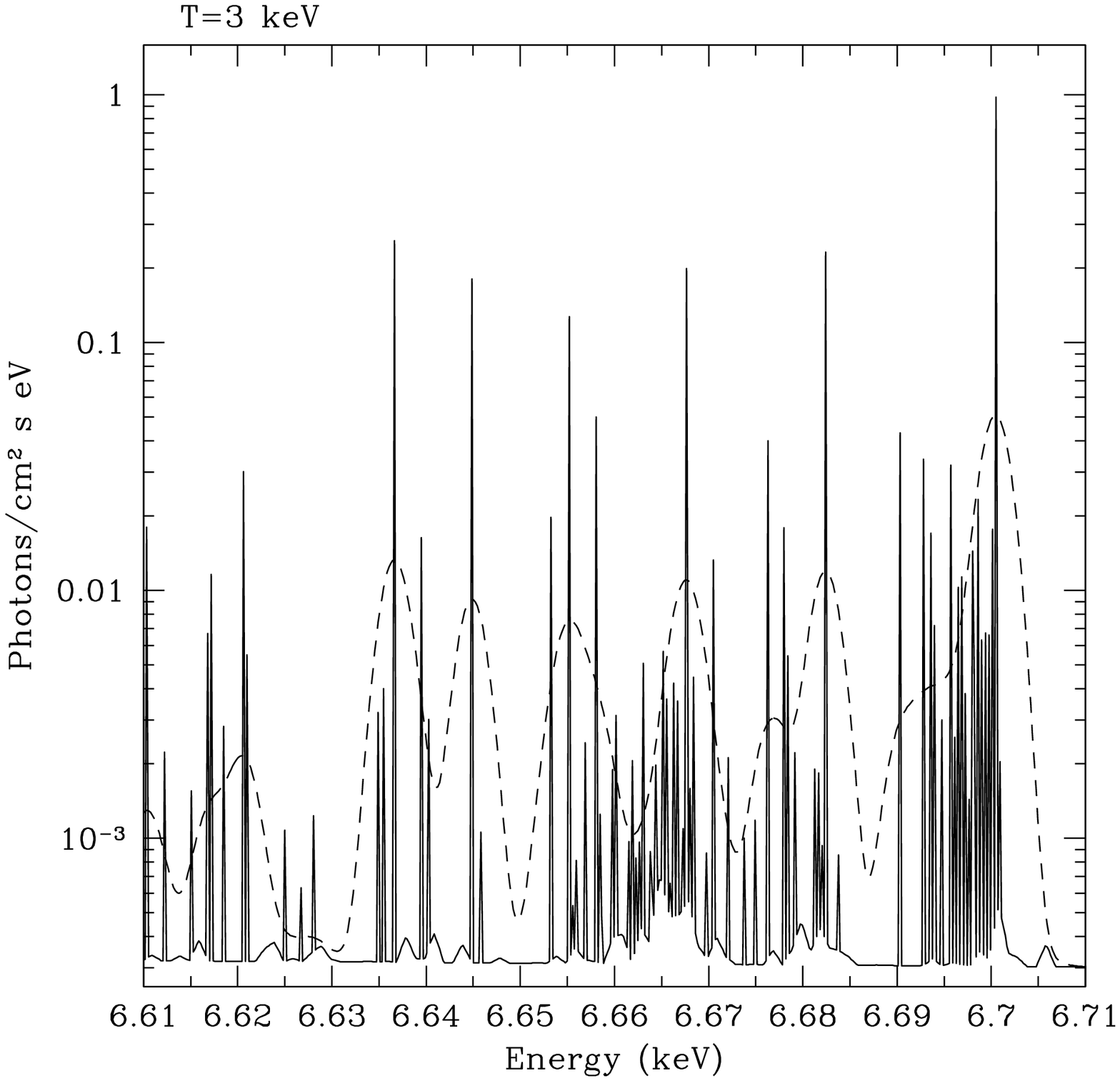}
\includegraphics[width=0.5\columnwidth]{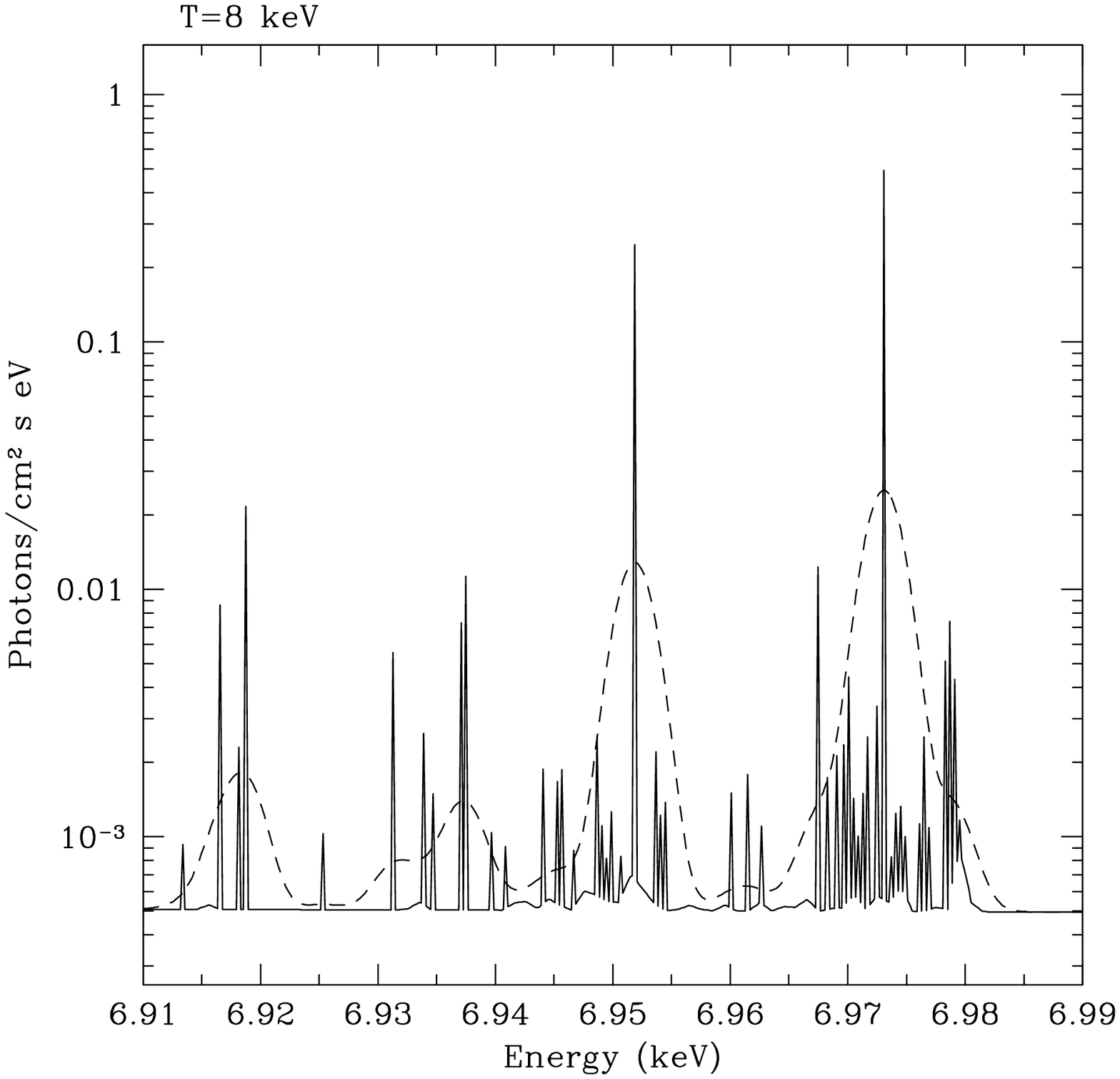}
}
\caption{
Spectra for the complexes of iron lines
near 6.7 keV at $kT_e$ = 3 keV {\bf (a)}
and near 6.9 keV at $kT_e$ = 8keV {\bf (b)}.
The dashed curves specify the thermal broadening of the lines.
The lines were normalized
to the most intense permitted helium-like iron $w$-line.
\label{fig:fig3}
}
\end{figure}

Fig. \ref{fig:fig3} (a) shows the radiation spectrum
for a hot plasma with a temperature of 3 keV
and a normal chemical abundance\footnote{
The abundance influences only the relative continuum intensity.
}
in the energy range\footnote{
The helium-like iron lines
and their satellites
are intense in this energy range.
}
near the permitted helium-like iron line at $\approx$ 6.7 keV.
The complex of iron lines
near the hydrogen-like iron line
($E_0 \approx$ 6.9 keV)
is shown in Fig. \ref{fig:fig3} (b)
at $kT$ = 8 keV.
These plots show the individual lines\footnote{
Without including their radiation widths.}
in the form of narrow peaks
with a width close to 10$^{-2}$ eV.
The plots were constructed by using the APEC code
(Astrophysical Plasma Emission Code; Smith et al. 2001),
which is part of the XSPEC V11.2 code.
All lines were normalized
to the most intense permitted $w$-line ($E_0$ = 6.7005 keV).
As an illustration,
the dashed curves in Fig. \ref{fig:fig3} (a) and (b)
indicate the same spectra
broadened by thermal motions of the ions.

\begin{figure}
\hbox{
\includegraphics[width=0.5\columnwidth]{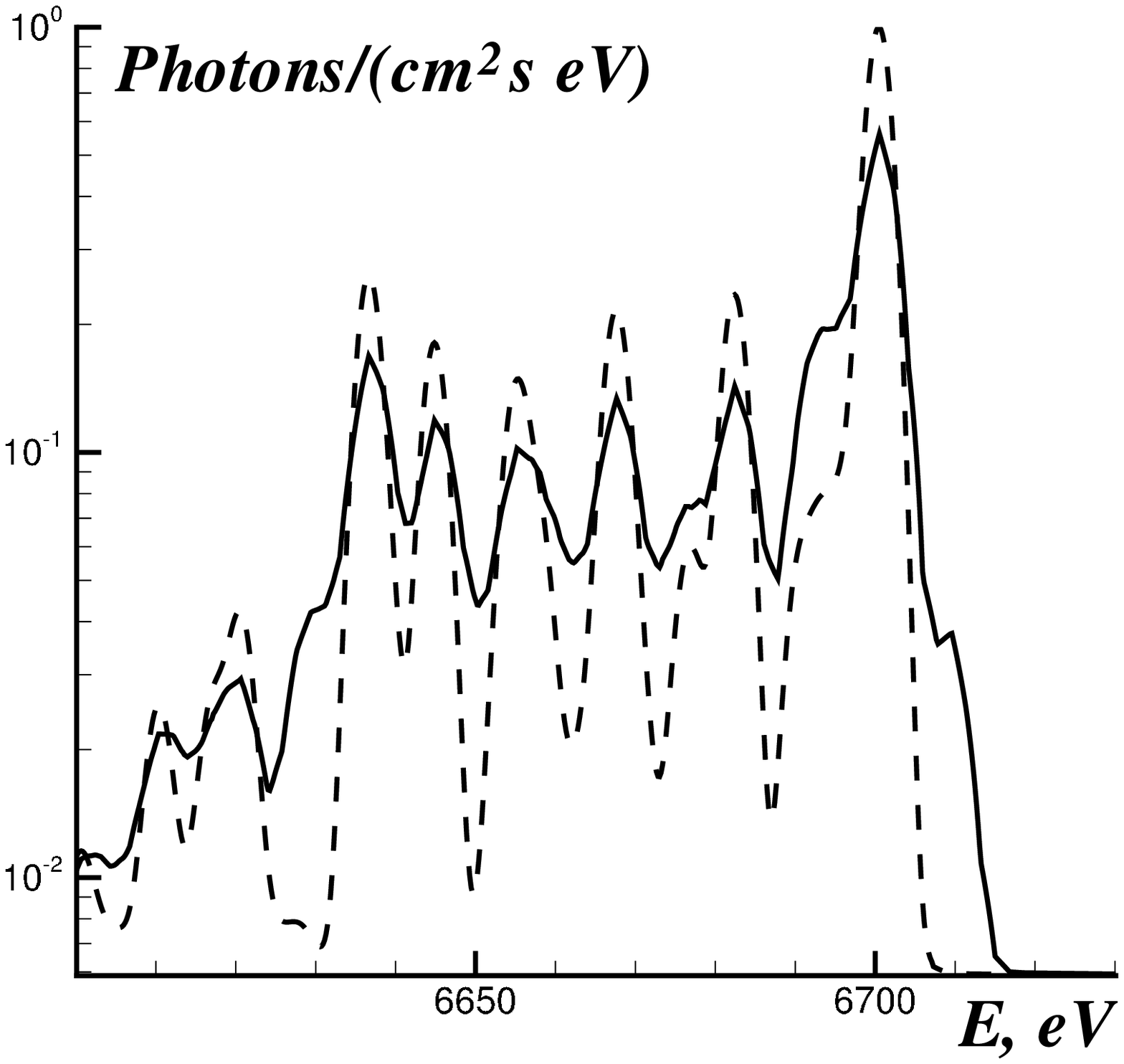}
\includegraphics[width=0.5\columnwidth]{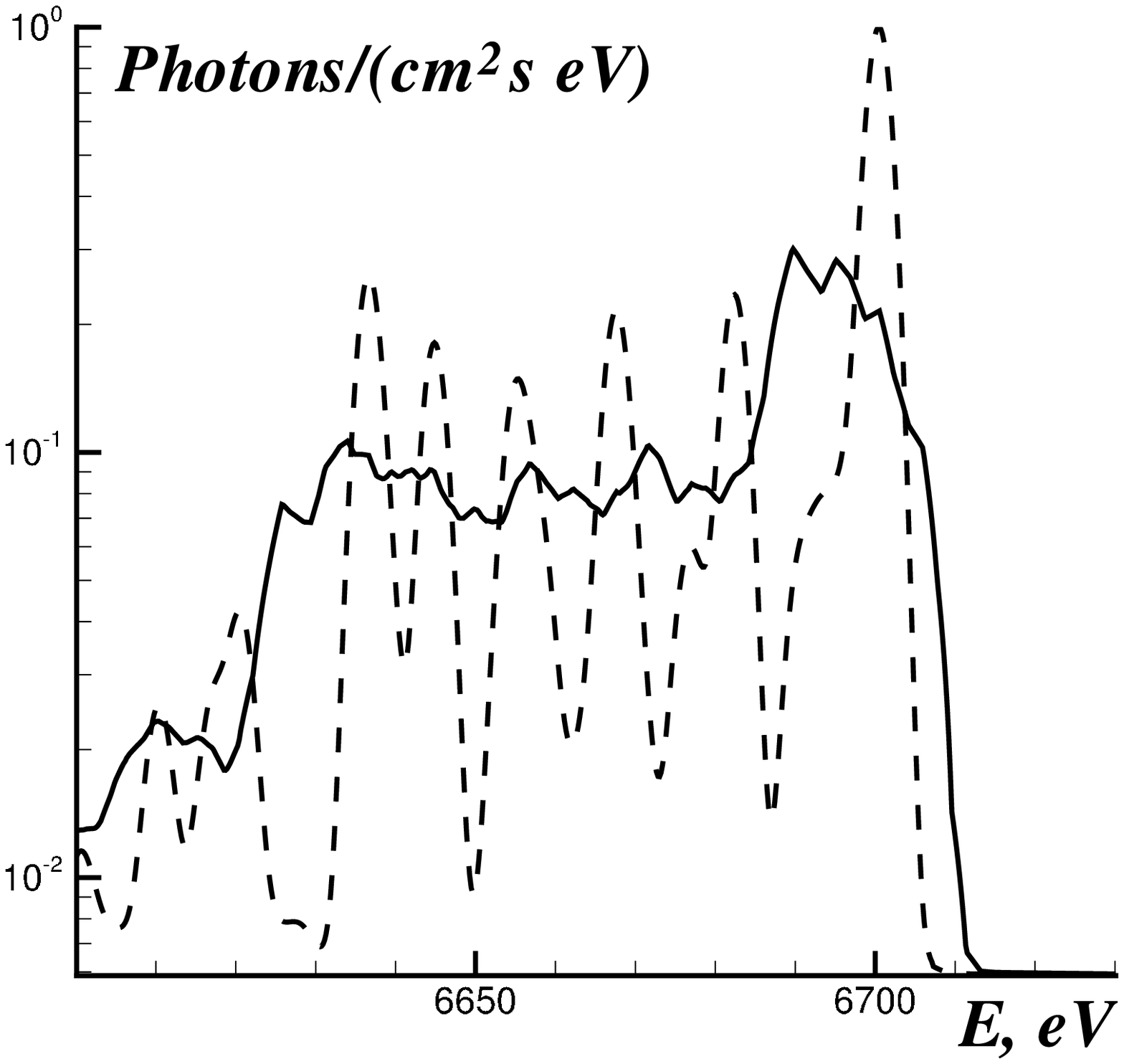}
}
\hbox{
\includegraphics[width=0.5\columnwidth]{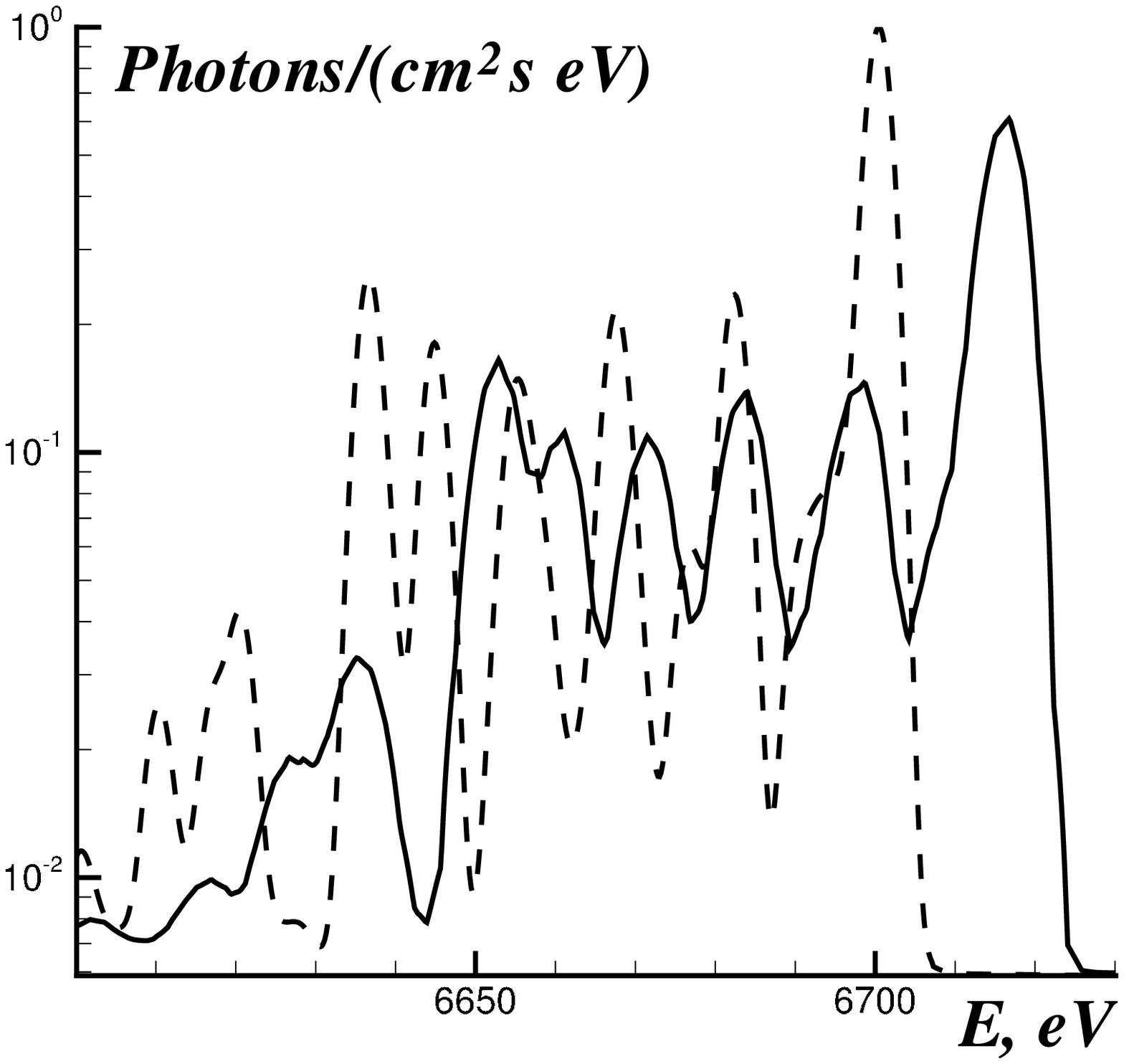}
\includegraphics[width=0.5\columnwidth]{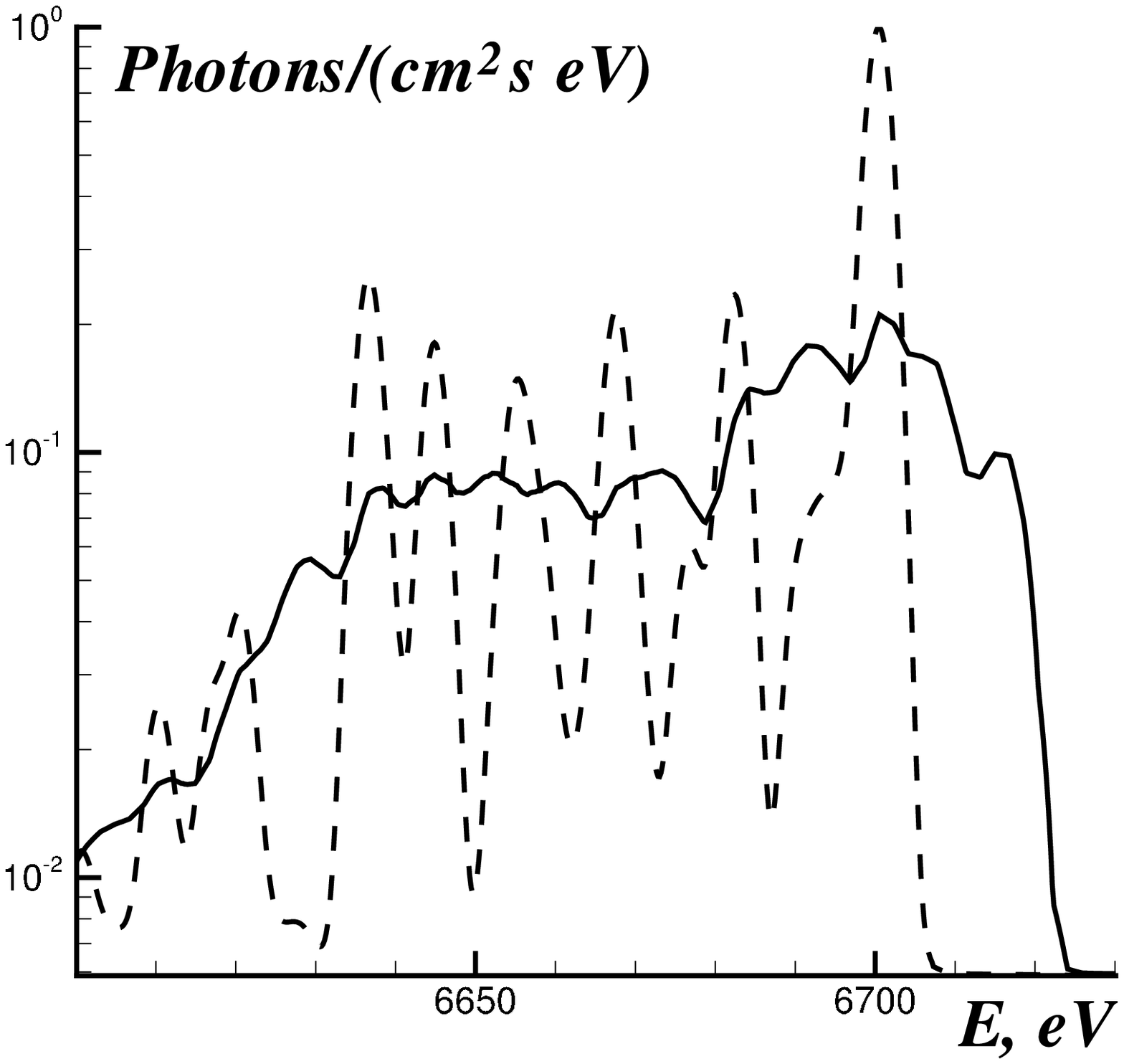}
}
\caption{
{\bf (a,b,c)} Spectra for the complex of iron lines
near 6.7 keV for $kT_e$ = 3 keV
are obtained
by folding with the profiles
shown in the upper middle, central, and lower left panels
of Fig. \ref{fig:fig2},
respectively.
The dashed curves
specify the spectrum
for the thermally broadened lines of the complex.
All curves
were normalized
to the same total number of photons in the complex of lines.
{\bf (d)} An approximate spectrum
for the complex of iron lines near 6.7 keV
from the cluster as a whole
obtained by folding the profile in Fig. \ref{fig:fig5}
with the spectrum in Fig. \ref{fig:fig3} (a).
\label{fig:fig4}
}
\end{figure}

Fig. \ref{fig:fig4} (a)--(c)
show the complex of lines near 6.7 keV
taking into account
both
thermal and hydrodynamic motions.
These spectra were computed by folding the profiles
presented in the upper middle, central, and lower left panels
of Fig. \ref{fig:fig2}
with the complex of individual lines and satellites
shown in Fig. \ref{fig:fig3} (a).
The dashed curves
again represent the broadening of the individual lines
due to thermal ion motions
in the absence of hydrodynamic motions.

The solid and dashed curves in Figs. \ref{fig:fig4} (a)--(c)
were normalized to the same number of photons
in the whole complex of lines.

The plasma emission
is a function of the temperature.
An increase in temperature
affects both
the relative line intensity
and the thermal line width.
At high temperatures (e.g., $kT$ = 8 keV),
the complex of hydrogen-like iron lines
near energy 6.9 keV
becomes intense,
Fig. \ref{fig:fig3} (b).

Using Figs. \ref{fig:fig3} (a) and (b), we wish to emphasize the following.
The most favorable energy range
for observing hydrodynamic iron line broadening effects
is to the right of the $w$-line center,
because there are virtually no other intense lines
at $kT$ = 3-10 keV
within several tens of eV of the center of this line.
Thus,
the far wings of the broadened line
can be studied in detail.

Of considerable interest is also the iron $z$-line
at photon energy $E_0$ = 6.6366 keV
that corresponds to the transition
from the 1s2s triplet to the ground state:
$^3S \rightarrow ^1S$.
In this case,
the left line wing
is favorable for observations.
It is possible to simulate the spectrum of the entire set of lines
in both complexes
and to compare it with the observed spectrum.

Fig. \ref{fig:fig4} (c) shows a very large shift of the line
observed from the column in the lower left panel in Fig. \ref{fig:fig2}.
All of this column flies toward us.
The velocity dispersion in it slightly exceeds
the thermal velocity dispersion.
In contrast,
the upper right panel of Fig. \ref{fig:fig2}
shows that the corresponding column flies away from us
with a slightly lower radial velocity
and larger velocity dispersion.
Collectively,
these two panels
suggest the presence of a large-scale motion
similar to the rotation
of the entire cluster gas
-- this is a manifestation of a large-scale mixer,
i.e., the last large merger.

In the central part of the cluster
(Figs. \ref{fig:fig2} and \ref{fig:fig4} (b)),
the broadening and the shift
are so large
that the individual lines and satellites in the complex of lines
are barely discernible.

\newpage

\vspace{0.4cm}

\small

\begin{figure}[htb]
\mbox{}\hfill\epsfxsize 0.6\hsize\epsfbox{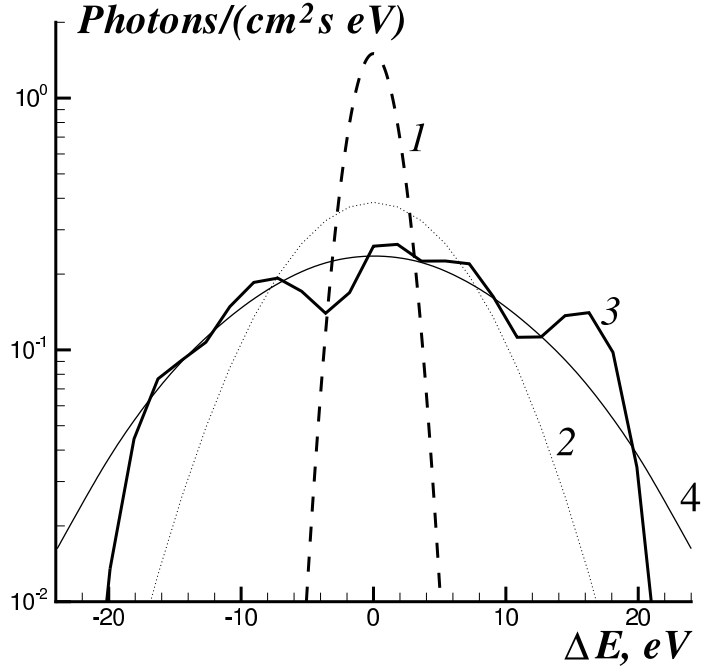}\hfill\mbox{}
\caption{
The iron line profile (curve 3) at $E_0=6.7$ keV
obtained by summing the nine profiles
shown in Fig. \ref{fig:fig2}.
This profile
was also normalized to the total number of photons.
To a first approximation,
this profile corresponds to the line profile
observed from the entire cluster.
The thermal broadening (curve 1)
and the two Gaussian curves
for the total broadening in the entire cluster
(total dispersion 460 km s$^{-1}$, curve 4)
and for dispersion $u_{\rm rms}$ = 270 km s$^{-1}$
(curve 2)
corresponding to the central panel
 in Fig. \ref{fig:fig2}
are shown for comparison.
\label{fig:fig5}
}
\end{figure}

\noindent {\bf Table 1.} Shifts and broadening of the X-ray line profile
at $E_0$ = 6.7 keV.
The positions of the cells in the table and the data in them
correspond to the spectra of the nine line-of-sight columns
presented in Fig. \ref{fig:fig2}.
The fourth row of the table
gives the velocity dispersion for the cluster as a whole.
This is dispersion of the integrated spectrum in Fig. \ref{fig:fig5}.
In the Table,
$u_1$ is the shift of the line profile center in [km s$^{-1}$],
and $u_{\rm HT}$
is the dispersion of the computed distributions from Fig. \ref{fig:fig2}.

\large

\begin{center} \begin{tabular}{|c|c|c|c|} \hline 
$u_1,\; u_{\rm HT} $&$+180,\;330$&$ -39,\;190 $&$-450,\;200 $\\ \hline
$u_1,\; u_{\rm HT} $&$+205,\;165$&$-205,\;280 $&$-470,\;190 $\\ \hline
$u_1,\; u_{\rm HT} $&$+695,\;110$&$+440,\;180 $&$ +80,\;140 $\\ \hline
$u_1,\; u_{\rm HT} $&$          $&$   0,\;460 $&$           $\\ \hline
\end{tabular}\end{center}

\normalsize

Since the ASTRO-E2 spectrometers have a limited angular resolution,
they will yield detailed images
only for the nearest rich clusters of galaxies.
For distant clusters,
the line profiles from the entire cluster
will be investigated.
Figs. \ref{fig:fig4} (d) and \ref{fig:fig5}
give the first idea of this profile.
A comparison of Fig. \ref{fig:fig4} (d)
with Figs. \ref{fig:fig4} (a)--(c)
shows
how informative
the set of high resolution spectroscopic data is
even with a limited angular resolution.
The profile in Fig. \ref{fig:fig5}
was obtained by averaging the nine profiles
presented in Fig. \ref{fig:fig2}
with equal statistical weights.
Recall that each of these nine profiles
was obtained for a very narrow unit column
in a cube with a volume of 160$^3$ mesh points.
Clearly,
the line profile from the entire cube
must be even smoother.
On the other hand,
the shifts of the profiles
shown in Fig. \ref{fig:fig2} and Table 1
suggest the presence of intense large-scale motions inside the
cluster.
In observations of the cluster as a whole,
they give a large contribution to the dispersion
but must leave traces in high-energy-resolution spectra.
The profile in Fig. \ref{fig:fig5}
is much broader
than any of the nine profiles in Fig. \ref{fig:fig2}
(see Table 1);
in particular,
it is broader than the profile in Fig. \ref{fig:fig1}.
This is the result of large-scale motions.

The spectrum in Fig. \ref{fig:fig4} (d)
was obtained by folding the profile in Fig. \ref{fig:fig5}
with the complex of iron lines near 6.7 keV
shown in Fig. \ref{fig:fig3} (a).
We see individual spectral features,
but their amplitude is appreciably smaller
than that
predicted in the model with thermal broadening
and a smaller hydrodynamic velocity dispersion.
The high-energy wing
of the permitted helium-like iron $w$-line
makes it possible
to judge the total velocity dispersion in the cluster.

The turbulent broadening
determined for all the nine profiles
shown in Fig. \ref{fig:fig2}
corresponds to a large
(at least severalfold)
decrease of the optical depth in the resonance $w$-line.
That is why we disregard the broadening of this line
due to resonant scattering (Gilfanov et al. 1987)
in the plots of Fig. 2--5.

\begin{figure}
\mbox{}\hfill\epsfxsize 0.6\hsize\epsfbox{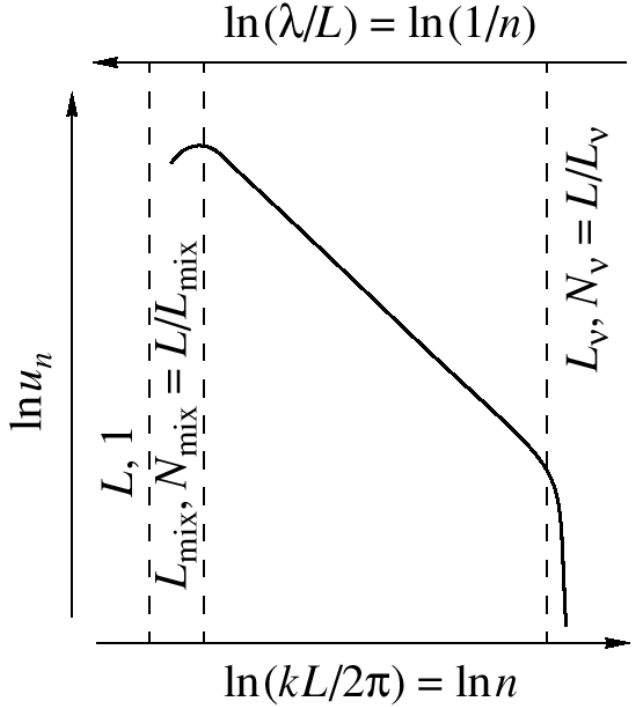}\hfill\mbox{}
\caption{
The $\lambda$ and $k$ fluctuation spectrum
in a Kolmogorov model.
The scales ($L$,1),
($L_{\rm mix}, N_{\rm mix}$),
and $(L_\nu, N_\nu)$
characterize the cluster,
the turbulence-maintaining mixers,
and the viscous dissipation region, respectively.
\label{fig:fig6}
}
\end{figure}

To elucidate the turbulence model,
Fig.\ref{fig:fig6} shows a Kolmogorov spectrum
of the Fourier velocity amplitudes $u_n$
with a power-law scaling
in the wavelength range
$L_\nu < \lambda < L_{\rm mix} \le L$
and harmonic numbers
$1 \le N_{\rm mix} < n < N_\nu$,
where$L$ is the cluster scale,
$\lambda = \lambda_n = L/n,\,$
$k = k_n = 2\pi/\lambda_n = 2\pi n/L,\,$
$ N_{\rm mix} = L/L_{\rm mix},\,$
$ N_\nu = L/L_\nu$,
and $L_\nu$ is the viscous damping scale.
Note that the total number of energy-containing eddies in volume $L^3$
is $N_{\rm mix}^3$.
Below,
we show
that the velocity dispersion
and the hydrodynamic Doppler broadening (1.2)
are determined by large-scale fluctuations
with $\lambda \sim  L_{\rm mix}$.
The fractal line profile fluctuations,
which are smoothed out by thermal broadening,
are associated with the Kolmogorov small-scale tail.

When computing the realization
shown in Figs. 1--5,
we took $N_{\rm mix}$ = 1 and $N_\nu$ = 40.
Because of the significant thermal broadening of the profile
and the decrease in small-scale pulsation velocity amplitude
with decreasing scale,
the difference between the realizations
for $N_\nu$ = 40 and $N_\nu \gg $ 40 is small
(see the full paper for a discussion).
The dependence on $N_{\rm mix}$ is discussed in Section 7 below.

We calculated the velocity field $u(x,y,z)$
in a cube $(2l)^3$
with an edge 2$l$.
The function $u(x,y,z)$
satisfies periodic boundary conditions
at the boundaries of the large cube $L^3$.
In our calculations (Figs. 1-5),
we took $L = 1$ and 2$l$ = 0.4.
We chose 2$l < L$
to reduce the influence of the periodic boundary conditions.
The cube $(2l)^3$
was covered by a 3D mesh of 160$^3$ computational points.
The separations between the mesh points are $\delta l = 2l/160$.
Each column in Fig. \ref{fig:fig2} has a length 2$l$
along the line of sight
and a cross section in the plane of the sky
in the form of a $\delta l \times \delta l$ square.

\subsection{Turbulence and the Kinematic SZ Effect}

The hydrodynamic velocity distribution of a cluster
in the plane of the sky
can also be analyzed by studying manifestations
of the kinematic SZ effect (Sunyaev and Zeldovich 1980),
i.e. by measuring the intensity fluctuations
of the cosmic microwave background radiation (CMB) within the cluster
$$
\frac{ \Delta T_r(y,z) }{T_r} = 
\sigma_T\,
\int_{-\infty}^{\infty}\,\frac{v_{||}(x,y,z)}{c}\, n_e(x,y,z)\,dx.
\eqno (1.4)
$$
Sunyaev et al. (2003) and Nagai et al. (2003)
used this method of analysis
in their numerical modeling.
Formula (1.4) was first suggested by Sunyaev and Zeldovich (1970)
to calculate the primordial Doppler CMB fluctuations
and was used by Sunyaev (1977)
in calculating the fluctuations due to secondary ionization.
In formula (1.4), $y$ and $z$
are the coordinates in the plane of the sky;
$x$ is the coordinate along the line of sight;
and $v_{||}$
is the component of the local hydrodynamic velocity vector
along the line of sight
(the $x$ velocity component)
below denoted by $u$,
$\hat v_{||} = v_{||}/u_{\rm rms};$
and $T_r$ is the CMB temperature.
The $\Delta T_r/T_r$ fluctuations (1.4)
differ only by a factor
from the fluctuations in the dimensionless total momentum
$$
p(y,z) = \int \, \hat n_e \, \hat v_{||}(x,y,z) \, dx/(2l),
\eqno (1.5)
$$
of the emitting matter along the $x$ axis
(the line-of-sight or column momentum),
where $\hat{n}_e = n_e/(n_e)_{\rm centr},$
$(n_e)_{\rm centr},$ is the central cluster density.
Below,
the density $n_e$ is assumed to be roughly uniform over the cluster core.

\begin{figure}
\includegraphics[angle=-90,width=\columnwidth]{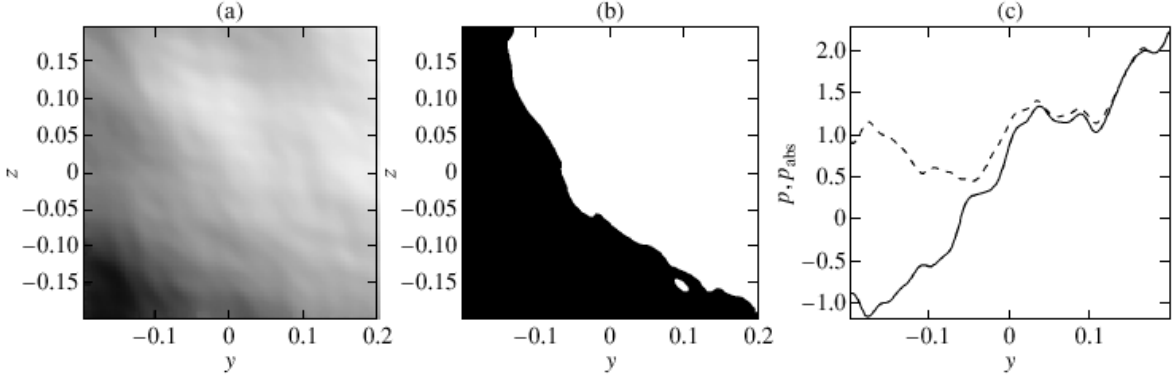}
\caption{
Distribution of the CMB intensity fluctuations (1.4)
in the $y,z$ plane of the sky
that follows from the Kolmogorov model.
{\bf (a)} A map of the line-of sight momentum $p(y,z);\,$ $L = 1$.
The white and black colors
correspond to the motion away from and toward the observer,
respectively.
The shades of gray
give the local values of
$\Delta T_r(y,z)/T_r.$
{\bf (b)} The $\Delta T_r(y,z)/T_r =0$ isoline
that separates the black and white regions
with opposite signs of $\Delta T_r.$
{\bf (c)} The $p(y,z=0)$ cut
(solid curve)
of the $p(y,z)$ surface.
Dashed line
represents the same cut
in the case of the function $p_{\rm abs}$
given by formula (1.6).
\label{fig:fig7}
}
\end{figure}

\begin{figure}
\mbox{}\hfill\epsfxsize 0.6\hsize\epsfbox{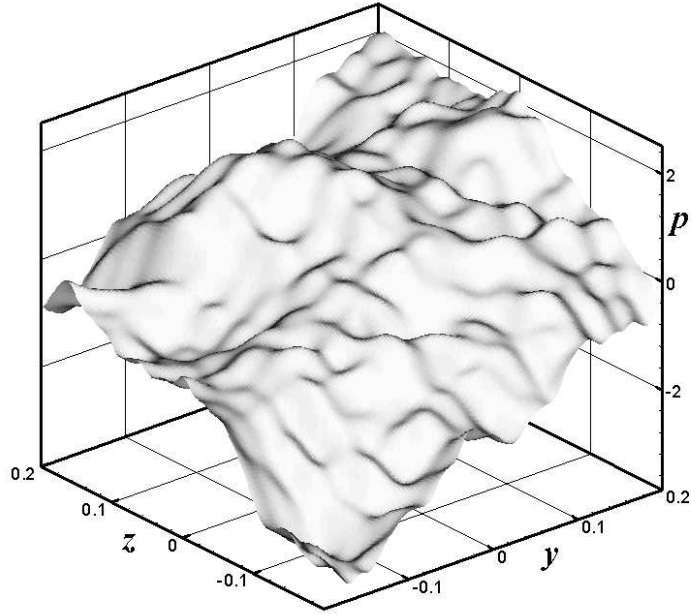}\hfill\mbox{}
\caption{
Relief of the kinematic SZ effect.
The height $p$
of the point $(y,z,p)$
on the $p(y,z)$ surface
(the height above the $(y, z)$ plane)
is proportional to the fluctuation amplitude
$\Delta T_r(y,z)/T_r$ (1.4).
\label{fig:fig8}
}
\end{figure}

Fig. \ref{fig:fig7} and \ref{fig:fig8} show maps, reliefs, and cuts
of the intensity fluctuations (1.4).
The degree of blackening in Fig. \ref{fig:fig7} (a)
is proportional to the local $\Delta T_r(y,z)/T_r$ fluctuation amplitude
and the local momentum $p(y,z)$.
Fig. \ref{fig:fig7} (b) shows regions in the $y, z$ plane of the sky
that move\footnote{
Here, we are dealing with the motion on average,
because $ p = \int \hat{u} \, dx/(2l) $
}
toward (black)
and away from us (white).
We performed our calculations
by using the Kolmogorov model
(for details, see the sections below).
For long-wavelength turbulence
with $N_{\rm mix} \sim 1$ (see Fig. \ref{fig:fig6}),
the size of the region with the same sign of $p$ (1.5)
is on the order of the cluster size $L$.
In the typical realization
shown in Figs. 1-5 and 7-10,
$N_{\rm mix}$ = 1 and $N_\nu$ = 40.
Fig. \ref{fig:fig7} (c) shows the cut of the $p(y,z)$ surface
(see Fig. \ref{fig:fig8})
by the $z = 0$ plane
that passes through the cluster center.
The solid and dashed curves in this figure
indicate, respectively, the functions $p(y,0)$ (1.5)
and
$$
p_{\rm abs}(y,0) = \int_{-l}^{l} \, \hat n_e \, |\hat u(x,y,0)|. \, dx/(2l)
\eqno (1.6)
$$
In the large-scale case ($N_{\rm mix} \sim 1$),
the functions $|p(y,z)|$ (1.5) and $p_{\rm abs} (y,z)$ (1.6)
are of the same order of magnitude,
as illustrated by Fig. \ref{fig:fig7} (c).

The sizes of the identically signed spots (Fig. \ref{fig:fig7} (b)
for the function $p(y,z)$
and the $|p|/p_{\rm abs}$ ratio
are determined by the scale $L_{\rm mix} = L/ N_{\rm mix}$.
If there are many mixers ($N_{\rm mix}\gg 1$),
then the function $p(y,z)$ changes sign
during a displacement $\Delta l_\perp \sim L_{\rm mix}$
in the plane of the sky
transverse to the line of sight.
In this case, $|p|/p_{\rm abs} \sim 1/\sqrt{N_{\rm mix}}$.
In deriving these estimates,
we assume that
(1) the separation between mixers is on the order of their size,
(2) the fluctuation amplitude of the velocity $u_{\rm rms}$
on each mixer
is of the same order of magnitude,
and (3) the velocity correlation
decays on a scale on the order of $L_{\rm mix}$.
In this case $p_{\rm abs}\sim 1$,
because
$\langle \hat u \rangle \sim 1,$
$\langle 2l     \rangle \sim 1$
(we set $\hat n_e$ = 1),
and $|p| \sim N^{-1/2}_{\rm mix}$,
because the total momentum $p(y,z)$ (1.5) along the $x$ line of sight
is the sum of the momenta of $N_{\rm mix}$ out-of-phase mixers.

It is interesting to calculate the Kolmogorov scaling law
for the CMB fluctuations
$\Delta T_r/T_r \propto p$ (1.4).
This law determines the amplitude of the fractal ``fringes''
(small fluctuations or spots)\footnote{
Small oscillations or fringes
are formed by a random addition of many eddies
with different scales;
see the sections below and the full paper.
}
in the dependences
$\Delta T_r(y,z)$ and $p(y,z)$
(see Figs. \ref{fig:fig7} (a), (c), and \ref{fig:fig8}).
The velocity is known
(see Sections 3 and 4)
to be given by
$$
\langle | v(r_2) - v(r_1) | \rangle =
(\Delta v)_{\Delta l} \propto
(\Delta l)^{1/3}, \;\;\;\; \Delta l = | r_2 - r_1|.
\eqno (1.7)
$$

Let us show that the following law holds for the kinematic SZ effect:
$$
\langle | \Delta T_r(r_{\bot 2}) - \Delta T_r(r_{\bot 1}) | \rangle
= [\Delta (\Delta T_r) ]_{\Delta l_{\bot}}
\propto (\Delta p)_{\Delta l_{\bot}}
\propto (\Delta l_{\bot})^{5/6},
\eqno (1.8)
$$
where
$
\Delta l_\bot = | r_{\bot 2} - r_{\bot 1} |,\,
$
$r_\bot = (y,z)$
is a two-dimensional vector
in the plane of sky.
We restrict our analysis to the case $N_{\rm mix} \sim 1$
presented in Figs. 1-5 and 7-10.
Below,
we give a brief derivation.
Let us consider two adjacent parallel $x$ lines of sight
separated by a distance $\Delta l_\bot$.
The value of $(\Delta p)_{\Delta l_\bot}$
is equal to the difference between the momenta $p$ (1.5)
on these lines of sight.
Let us calculate this difference.
The contributions of large eddies
with a size $l_{\rm rot}\,$
$(l_{\rm rot} > \Delta l_\bot)$
to the difference between integrals (1.5) are small,
because these contributions
are almost equal on adjacent lines of sight.
The contribution of small eddies
$(l_{\rm rot} < \Delta l_\bot)$
increases with size $l_{\rm rot}$
and reaches a maximum at $(l_{\rm rot} \sim \Delta l_\bot)$.
Consequently,
eddies with sizes on the order of the separation
between the lines of sight
should be considered to estimate (1.8).

These eddies are uncorrelated.
Indeed,
the correlation in the velocity field
induced by them
decays on a scale $\sim \Delta l_\bot$.
Therefore,
the corresponding contribution to the difference
$\Delta ( \Delta T_r ) \propto \Delta p$
is accumulated through a random addition.
According to formulas (1.5) and (1.7),
the momentum from a single eddy is
$$
\sim u_{\rm rms} \,
(\Delta l_\bot/L_{\rm mix})^{-1/3}\,
(\Delta l_\bot/L)/u_{\rm rms}.
$$
There are $\sim L/\Delta l_\bot$ such eddies
on a line of sight
which has a length $\sim L \sim L_{\rm mix}$.
Expression (1.8) follows from this
and from the fact
that the effect
is proportional to the square root
of number of small-scale eddies under consideration,
because
\[
(\Delta p)_{\Delta l_{\bot}} \sim
( \Delta l_{\bot}/L_{\rm mix})^{-1/3}\,
( \Delta l_{\bot}/L )\,
\sqrt{ L/\Delta l_{\bot} }.
\]
Scaling (1.8) is valid in the Kolmogorov range of scales
$L_\nu < \Delta l_{\bot} < L_{\rm mix}$.
In observations of $\Delta T_r/T_r \propto p$
or in direct numerical simulations,
the small-scale limit
is determined by the aperture resolution
or the spatial mesh step,
respectively.

\subsection{Comparison of the Two Methods for Studying Turbulence}

Let us compare the two effects under discussion,
the Doppler broadening of X-ray lines
and the kinematic SZ effect.
The map of CMB intensity (1.4)
carries information about the momentum
of a unit column along the line of sight.
This gives the velocity
relative to the frame of reference
associated with the CMB near the cluster\footnote{
The velocity of the cluster as a whole
can be determined by integrating the line profile
over the plane of the sky
within the cluster
and finding its centroid.
This operation will allow us
to measure the sum of the cluster recession velocity
due to the expansion of the Universe (redshift)
and the line-of-sight peculiar cluster velocity
in the frame of reference
associated with the CMB.
The kinematic SZ effect
allows this peculiar velocity component
to be determined.
},
i.e. its peculiar velocity.
The X-ray line profiles are more informative.
In the simplest case of an isothermal plasma
with a constant iron abundance,
they provide information
about the velocity distribution of matter
relative to the velocity of the cluster as a whole
(in particular, the velocity dispersion).
Below,
we discuss in detail several characteristic features of the spectrum
and their relationship to the turbulent velocity field.

\begin{figure}
\mbox{}\hfill\epsfxsize 0.9\hsize\epsfbox{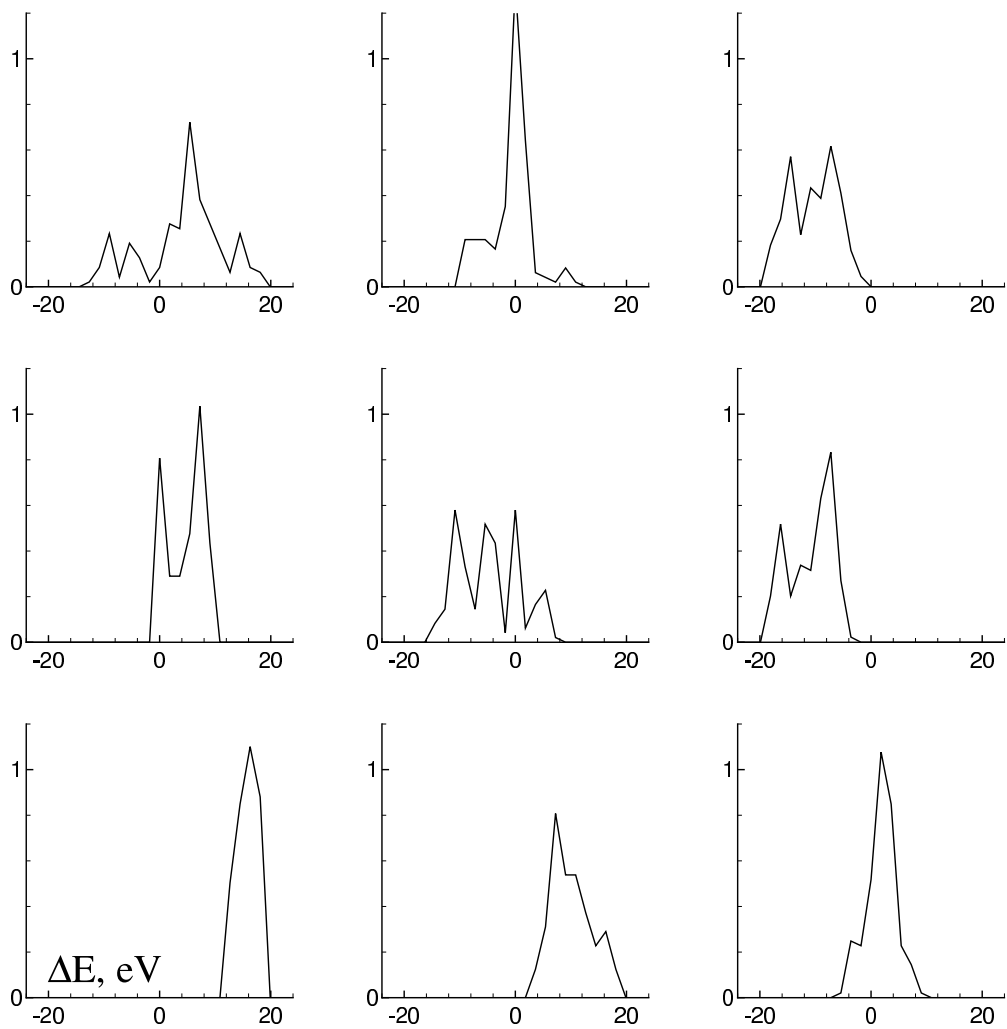}\hfill\mbox{}
\caption{
A map of the $f_H(\Delta E,y,z)$ profiles
for iron ions in the $(y,z)$ plane
at the points that coincide with the center,
vertices, and middles of the sides of the square
shown in Fig. \ref{fig:fig7} (a).
\label{fig:fig9}
}
\end{figure}

To compare the two effects,
Fig. \ref{fig:fig9} shows a map of the velocity distributions
$f_H (\Delta E)$
over the $(y,z)$ cluster plane.
The distributions $f_H (\Delta E)$
were computed at the points
located at the center, vertices, and middles of the sides
of the $2l \times 2l$ square
shown in Figs. \ref{fig:fig7} and \ref{fig:fig8}.
We took a frame of reference
in which the total momentum $\int \int p(y,z)dydz$
of the emitting and scattering matter
is equal to zero.
We are looking for the distributions
$f_H(\hat{u})$
in this frame.
The scale of dimensionless velocities $\hat{u}$
was recalculated to energy shifts $\Delta E$
by using formula (1.3)
at $u_{\rm rms} = 270$ km s$^{-1}$.
The velocity dispersion
changes from point to point in the plane of the sky,
$u_{\rm rms} = u_{\rm rms}(y,z)$.
For definiteness,
in the normalization $u_{\rm rms} = 270$ km s$^{-1}$,
we chose the dispersion at the center $y = 0, z = 0$.
Note that the dispersion over the cluster as a whole
appreciably\footnote{
by a factor of 1.5 to 2.5; cf. parabolas 2 and 4 in Fig. \ref{fig:fig5}.}
exceeds the local dispersion.
The relief of the kinematic SZ effect
is determined by the shift velocity distribution
(average velocity of the column as a whole).
We see from Figs. \ref{fig:fig2} and \ref{fig:fig5}
that these shift velocities
give a large contribution
to the velocity dispersion in the cluster as a whole.

\begin{figure}
\mbox{}\hfill\epsfxsize 0.8\hsize\epsfbox{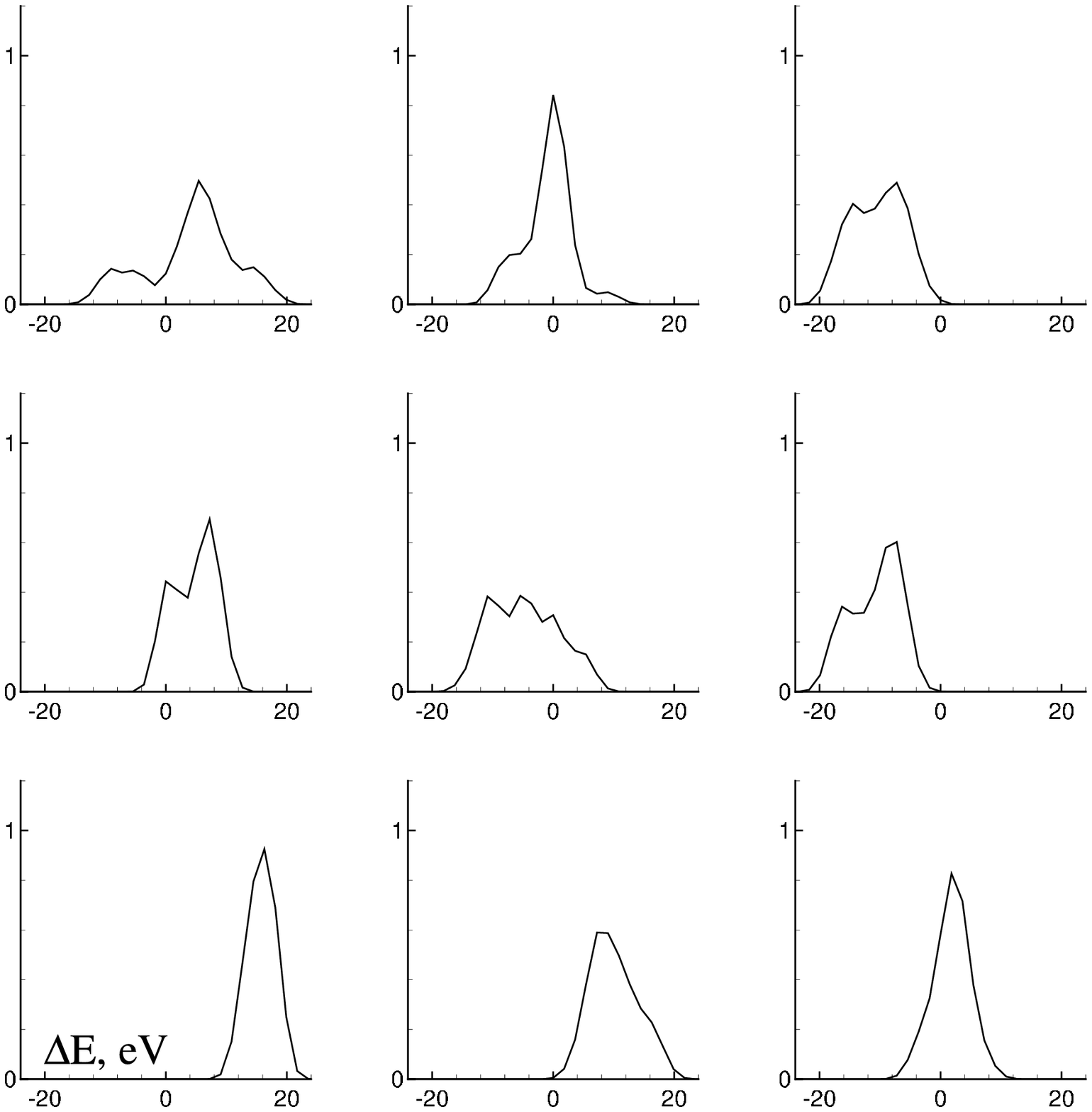}\hfill\mbox{}
\caption{
The $f_{\rm HT}(\Delta E)$ profiles
of Fe XXV lines at $kT$ = 3 keV
(the same as Fig. \ref{fig:fig2} but in linear coordinates).
The spectrum
shown in the middle panel of the upper row
from the three panels
was used to construct Figs. \ref{fig:fig1} and \ref{fig:fig4} (a).
The central and bottom left panels
were used in Figs. \ref{fig:fig4} (b) and (c).
\label{fig:fig10}
}
\end{figure}

The $f_{\rm HT}(\Delta E)$ profiles
of Fe XXV $(E_0 = 6.7 keV)$ lines
with turbulent and thermal broadenings
are drawn in Fig. \ref{fig:fig10}.
They were obtained
from the distributions $f_H(\Delta E)$ in Fig. \ref{fig:fig9}
by folding with the Maxwellian distribution at $kT = 3$ keV.
Thermal smoothing ``blurs'' some of the features
in the hydrodynamic distribution $f_H$\footnote{
At $u_{\rm rms} = 270$ km s$^{-1}$
and $kT = 3$ keV,
$\Delta \nu_{\rm DH}/\Delta \nu_{\rm DT} = 2.7.$
},
although the spectra remain fairly complex
and varied.
The centers
of the $f_H(\Delta E)$ and $f_{\rm HT}(\Delta E)$ profiles
coincide.
Of course, they coincide with the profile center
determined from the line-of-sight momentum $p(y,z)$.
Therefore,
the map of the kinematic SZ effect (Figs. \ref{fig:fig7} and \ref{fig:fig8})
is simultaneously the map of the shifts
in the profile centers in Figs. \ref{fig:fig9} and \ref{fig:fig10}.
Let us show this by using an example.
The black color in the left panel of Fig. \ref{fig:fig7} (a)
(the lower left corner of the square)
corresponds to high velocities
directed toward the observer.
These velocities clearly show up
in the rightward shift of the profile
along the energy axis
from $\Delta E = 0$
in the spectrum in the lower left corner of Figs. \ref{fig:fig9}
 and \ref{fig:fig10},
because motion toward the observer causes the photon energy to change.
The shifts of the profile center
as the position varies in the plane of the sky
are large
(3-6 local velocity dispersions).
Thus, it is possible to independently and experimentally determine
the shifts of the profile center
by two different methods
(the kinematic SZ effect and analysis of the line profile).

The line shape completely changes
during the shift to a large distance
$\Delta l_{\perp} \sim L$ in a direction
transverse to the $x$ line of sight
$(N_{\rm mix} \sim 1)$.
The shape changes continuously:
the change in shape is small for small shifts.
For example,
the variation of such a parameter
as the shift of the profile center during transverse shifts
is shown in Fig. \ref{fig:fig7} (c).
The profiles significantly deviate from the Gaussian profiles
(a large shift of the spectrum center,
splitting of the maximum,
a sharp break in the exponential tails).
A comparison with the Gaussian distribution
for the middle spectrum from the upper triplet of spectra
(Figs. \ref{fig:fig2} and \ref{fig:fig10})
is shown in Fig. \ref{fig:fig1}.

The presence of turbulence in the intergalactic cluster gas
and its amplitude
can also be judged
from differences in the intensity distributions
of permitted
from one side
and forbidden, intercombination lines
and their satellites
from another side
over the cluster (Gilfanov et al. 1987).
The optical depth in permitted lines
can appreciably exceed unity
in the absence of turbulence.
Turbulent broadening
reduces the optical depth in lines
and this effect.
{\it This is a third independent method
for studying turbulence in clusters of galaxies}
(Churazov et al., 2003).

Note that Fabian et al. (2003)
pointed to the existence of long cold $H_{\alpha}$-emitting filaments
in the Perseus cluster of galaxies
as an argument against the well-developed turbulence in this cluster.

The full paper is arranged as follows.
Section 2 gives a geometrical formulation of the problem
with projection onto the observer's direction (line of sight).
Section 3 presents three main components of the Kolmogorov model:
two boundaries with the sizes $(L_{\rm mix},$ $L_{\nu})$
of the inertial hierarchy of eddies
and fluctuation amplitude scaling.
Section 4 is devoted to the Fourier expansion
with random phases of the turbulent velocity field.
The procedure for calculating the velocity distribution
$f_H(u)$
from the specific realization $u(x)$ of a random velocity field
is described in Section 5.
The distribution $f_H(u)$
can be approximately calculated from the first moments
of the turbulent velocity field (Section 6).
When there are many mixers, $N_{\rm mix}\gg 1,$
the turbulent distribution $f_H(u)$
tends to a Gaussian distribution $f^G(u)$
(Section 7).
The superposition of thermal and hydrodynamic broadenings
is analyzed in Section 8.
A theory of the turbulent profile shape
is constructed in Sections 9 and 10.
We elucidate the question of typical spikes in the profile
and its wings
and the variety of possible shapes.
The transverse correlations
(in the plane of the sky)
are studied in Section 11.
The Lorentz line broadening is taken into account in Sections 12 and 13.

\section{Turbulence and Doppler Shift}

We are interested in the emission
from the hot intergalactic plasma
that fills cluster of galaxies.
The plasma is in hydrostatic equilibrium in the gravitational well
produced by collision-less dark matter.
The density decreases toward the periphery of the well
on scales of the order of the cluster scale L.
The kinetic energy of the turbulent motion
caused by mergers
reaches $\sim 15 \%$ of the thermal energy.
By the meaning of the formulated problem
(Doppler shift),
the hydrodynamic density and plasma temperature fluctuations,
which affect the local intensity
but not the frequency,
are of little importance.
The velocity fluctuations are dominant.

There is a three-dimensional velocity field $\vec v(\vec r, t)$.
It is necessary to determine the Doppler shift in a frame of reference
in which the cluster as a whole is at rest.
We will consider the line profile
when the X-ray telescope
is pointed at a point in the cluster plane of the sky.
The profile is defined by the function $u(x,t)$
that specifies the velocity $u$
along the $x$ line of sight
passing through the observer
and the point
at which the telescope is pointed.
The small eddies that are localized at the points
separated along the $x$ axis
are statistically equivalent.
The turnover time of large-scale eddies
is much longer
than the time
it takes for light
to pass through the cluster.
In addition, the evolution time
of even the smallest observed velocity fluctuations
is much longer
than the exposure time
during which an object is observed from a satellite.
Consequently,
we may omit the time dependence of the function $u(x,t)$
and deal with the instantaneous velocity field $u(x)$.

The Doppler change in photon energy $\Delta E = E - E_0$ is
$$
\Delta E/E_0 = - u/c, \;\;\;\;\;
\Delta E\,[\,{\rm eV}\,]=
6.7\,
(u/300\,[\,{\rm km s}^{-1}\,]\,)\,
(E_0/6700\,[\,{\rm eV}\,]\,)
\eqno (2.1)
$$
where $E_0$ is the energy of the photon
emitted from a region
that is at rest
relative to the cluster as a whole,
and E is the energy of the photon
emerged from a region
that moves with velocity $u$
relative to the adopted frame of reference.
Let the cluster be to the right of the observer on the $x$ axis.
Accordingly,
the velocity $u$ is positive
if the emitting region recedes from the observer.

\section{The Kolmogorov Model}

An individual line profile
corresponds to each velocity field $u(x)$.
We are interested in statistically representative realizations of
$u(x)$.
The Kolmogorov views of turbulence
are of great importance in constructing a typical profile.
The point is that
they solve the difficult question
regarding the contribution of small-scale fluctuations.

In a Kolmogorov cascade,
energy is transferred from large to small scales.
The largest eddies have scales on the order of the mixer size
$L_{\rm mix}$.
The size $L_{\nu}$ of the smallest eddies
is determined by viscous dissipation.
A cluster of volume $L^3$
is assumed to be covered by a three-dimensional mesh
of $N^3_{\rm mix}$ statistically equivalent mixers,
$$
N_{\rm mix} = L/L_{\rm mix}. \eqno (3.1)
$$
The $\vec v(\vec r)$ fluctuations
are isotropic in velocity vector orientation.
The $x$ axis is not highlighted in any way.
Therefore, the Kolmogorov scaling (1.7)
can be written for the $x$ velocity component
(the projection of three-dimensional fluctuations
onto the line of sight).
In the hierarchy of scales,
the mean energy losses $\dot E$
within the inertial range
are scale-independent (Kolmogorov 1941),
$$
\dot E \sim [\,(\Delta u)_{\Delta t}\,]^2 / \Delta t
\sim [ \, (\Delta u)_{\Delta x} \, ]^3 / \Delta x,
\eqno (3.2)
$$
where $\dot{E}$
are the hydrodynamic energy losses per gram of matter,
and $\Delta u$ is the velocity variation
due to the shift in time $\Delta t$
or along the line of sight $\Delta x$.
The Kolmogorov scaling follows from (3.2):
$$
(\Delta u)_{\Delta x} \sim \dot E^{1/3} \, (\Delta x)^{1/3}.
\eqno (3.3)
$$

The locally viscous dissipation is given by the formula
$$
\dot E =
(\nu/2)
 \sum \sum\,
 ( \partial v_i/\partial x_j + \partial v_j/\partial x_i )^2
\sim \nu\, [\, (\Delta u)_{\nu} \,]^2/[ \, (\Delta x)_{\nu} \, ]^2, 
\eqno (3.4)
$$
where $(\Delta u)_{\nu}$ and $(\Delta x)_{\nu}$ are, respectively,
the velocity and diameter scales of eddies
on which viscous dissipation takes place.
To estimate these scales,
law (3.3) is extended to the range of dissipation through viscosity
(Kolmogorov 1941; Monin and Yanglom 1965; Landau and Lifshitz 1986).
We then find from formula (3.4) that
$$
(\Delta x)_{\nu} \sim
 \nu^{3/4}/\dot E^{1/4} \sim
 (c_s/u_{\rm rms})^{3/4} \,
 l_p^{3/4} \,
 L_{\rm mix}^{1/4},
\eqno (3.5)
$$ 
$$
N_{\nu} = L/ (\Delta x)_{\nu}, \;\;\;\;\;
(\Delta t)_{\nu} \sim
 \sqrt{\nu/\dot E},
\eqno (3.6)
$$
where $c_s$ is the speed of sound,
$u_{\rm rms}$ is the characteristic hydrodynamic velocity
(rms deviation or dispersion)
defined by the second moment
$$
(\, u_{\rm rms} \,)^2 = \int u^2\, dx/ L
$$
of the function $u(x)$
(the integral is taken over a range on the order of $L$),
and $l_p$ is the proton mean free path.
Large-scale eddies mainly contribute to the dispersion
$u_{\rm rms}$
and the kinetic energy of the turbulence.
$N_{\nu}$ (3.5) gives the number of the smallest eddies
on the characteristic cluster scale.
Formulas (3.5) and (3.6) define the smallest scales.

The size of small eddies (3.5)
is mainly determined by the proton mean free path $l_p$.
The latter varies over a wide range
(many orders of magnitude)
with amplitude of the random magnetic field
in the intergalactic plasma of the cluster.
Below,
we give the corresponding estimates.
In the absence of a magnetic field,
the plasma viscosity
$$
\nu = 0.48 \, v_{\rm Tp} \, l_p =
1.5 \cdot 10^{27} \, 
( k T )^{5/2} / n_{-2},
$$
$$
l_p =
v_{\rm Tp} \, \tau_p =
7.2 \cdot 10^{19} \, ( kT )^2 / n_{-2}, \;\;\;\;
v_{\rm Tp} = \sqrt{2 k T / m_p }
$$
is determined by Coulomb collisions between ions
(see, e.g., Rosenbluth and Sagdeev 1983);
in what follows,
the kinematic viscosity $\nu$
and the mean free path $l_p$ are in CGS units,
the temperature kT is in keV,
$
n = 10^{-2} n_{-2}
$ cm$^{-3}$,
$m_p$ is the proton mass.
The gyroviscosity of plasma with a magnetic field
is given by the expression (Rosenbluth and Sagdeev 1983)
$$
\nu = 0.5 \,
kT/m_p \, \omega_{\rm cp}
= 0.25 \, v_{\rm Tp} \, r_L,\;\;\;\;\;
r_L = m_p \, v_{\rm Tp} \, c/eH,
$$
where $\omega_{\rm cp}$ is the cyclotron frequency of protons,
and $r_L$ is their Larmor radius.
The viscous scale ratio is
$$
\frac{\Delta x_{\nu}(H=0)}{\Delta x_{\nu} (H \neq 0)} =
(1.92 \, \tau_p \, \omega_{\rm cp})^{3/4} =
2.3 \cdot 10^6 \, (\,H_{-8}\,)^{3/4} \,
(\, kT \,)^{9/8} \, (\, n_{-2} \,)^{-3/4},
$$
where $H = 10^{-8} H_{-8}$ G.
Accordingly,
the dimensionless viscous wave numbers $N_{\nu}$ (3.5) are
\[
 N_{\nu}(H=0) = 900 \, (\, N_{\rm mix} \,)^{1/4} \,
 \Ma^{3/4} \,
(\, L_{200} \,)^{3/4} \,
 (\, n_{-2} \,)^{3/4} \,
(\, kT \,)^{ - 3/2},
\]
\[
N_{\nu}( H \neq 0) =
 3.5 \cdot 10^9 \,
(\, N_{\rm mix} \,)^{1/4} \,
 \Ma^{3/4} \,
(\, L_{200} \,)^{3/4} \,
(\, H_{-8} \,)^{3/4} \,
(\, kT \,)^{ - 3/8},
\]
where Ma = $u_{\rm rms}/c_s;$
$u_{\rm rms}$ and $c_s$
are the characteristic hydrodynamic velocity and the speed of sound,
respectively;
and $L = 200 L_{200}$ kpc.

In a hot rarefied plasma without a magnetic field,
the mean free path $l_p$ is very large.
Thus, a situation where $N_{\nu}$ is very small
$( N_{\nu} \sim 1 )$
is hypothetically possible.
As we show in Sections 9 and 10 of the full paper,
the line profile in this situation exhibits features
that allow it to be distinguished
from the case with $N_{\nu} \gg 1$.
Currently available estimates
based on Faraday rotation measurements
give magnetic field strengths in the range 1-10 $\mu$G
(see, e.g., Ge and Owen 1993),
suggesting that the Kolmogorov range
in the intergalactic turbulence spectrum
is very wide.

\section{Fourier Expansion}

For the line profile to be constructed,
we must know the velocity distribution $f_H(u)$.
The distribution $f_H(u)$
can be calculated from the velocity field $u(x)$.
Let us write out an analytical formula for the function $u(x)$
that satisfies law (3.3).
It would be natural
to represent the stochastic turbulent motions
in the Kolmogorov range of scales as a Fourier series
$$
u'(x; \vec \psi) =
(\, N_{\rm mix} \,)^{\alpha}
\, {\rm Re }
 \sum_{ n = N_{\rm mix} }^{N_{\nu}}
 n^{ - \alpha - 1/2 } \,
 \exp( i \psi_n ) \,
\exp \left( i 2\pi n x / L \right).
\eqno (4.1)
$$
In formula (4.1), the index $\alpha$
specifies the velocity fluctuation scaling
$(\Delta u)_{\Delta x} \propto (\Delta x)^{\alpha}$.
In case (3.3), $\alpha = 1/3$.
The phases $\vec \psi = \{ \psi_n \}$ in the harmonic expansion (4.1)
are independent random variables
uniformly distributed in the segment [0,2$\pi$].
The velocity
is given by the real part of expansion (4.1) (the symbol Re).
The summation limits (4.1) coincide with boundaries (3.1) and (3.5)
of the inertial range.
The function of $x$ (4.1)
is considered in the segment $[l_1, l_2]$
with a length on the order of $L$.
Series (4.1) describes the fractal curve $u(x)$
whose fluctuations satisfy the Kolmogorov scaling law (3.3).
This series makes it possible to carry out specific calculations.

Let us ``displace'' the auxiliary function u' (4.1)
so that the velocity field is centered $(\int u_0 \, d_x = 0)$.
We change $u'\rightarrow u_0$, where
$$
u_0(x) = u'(x) - u_1,\;\,
u_1 = \int_{l_1}^{l_2} u'(x)\, dx/l_{21},\;\;
l_{21} = l_2 - l_1, \; (l_1, l_2) \sim L, \eqno (4.2)
$$
the function $u'$ is calculated from formula (4.1).
We normalize the function $u_0$ (4.2)
to the rms velocity $u_{\rm rms}$.
By definition,
$u_{rms}$ gives the dispersion $\sigma$ of the velocity distribution.
The velocity dispersion,
along with the kinetic energy of the turbulence,
is determined by large eddies.
We write
$$
u(x) = u_0(x)/u_{\rm rms},\;\;\;\;
u_{\rm rms}^2 = \int_{l_1}^{l_2} [\, u_0(x) \,]^2 \, dx/l_{21}. \eqno (4.3)
$$
Below, we deal with the centered (4.2)
and normalized (4.3) function $u(x)$
defined by formulas (4.1)-(4.3).
The model of a homogeneous cloud with size $L$
is used as the first approximation to the density distribution.
Clearly, this picture can be easily generalized
to the real density distribution in clusters of galaxies.

\section{The Velocity Distribution}

\begin{figure}
\mbox{}\hfill\epsfxsize 1\hsize\epsfbox{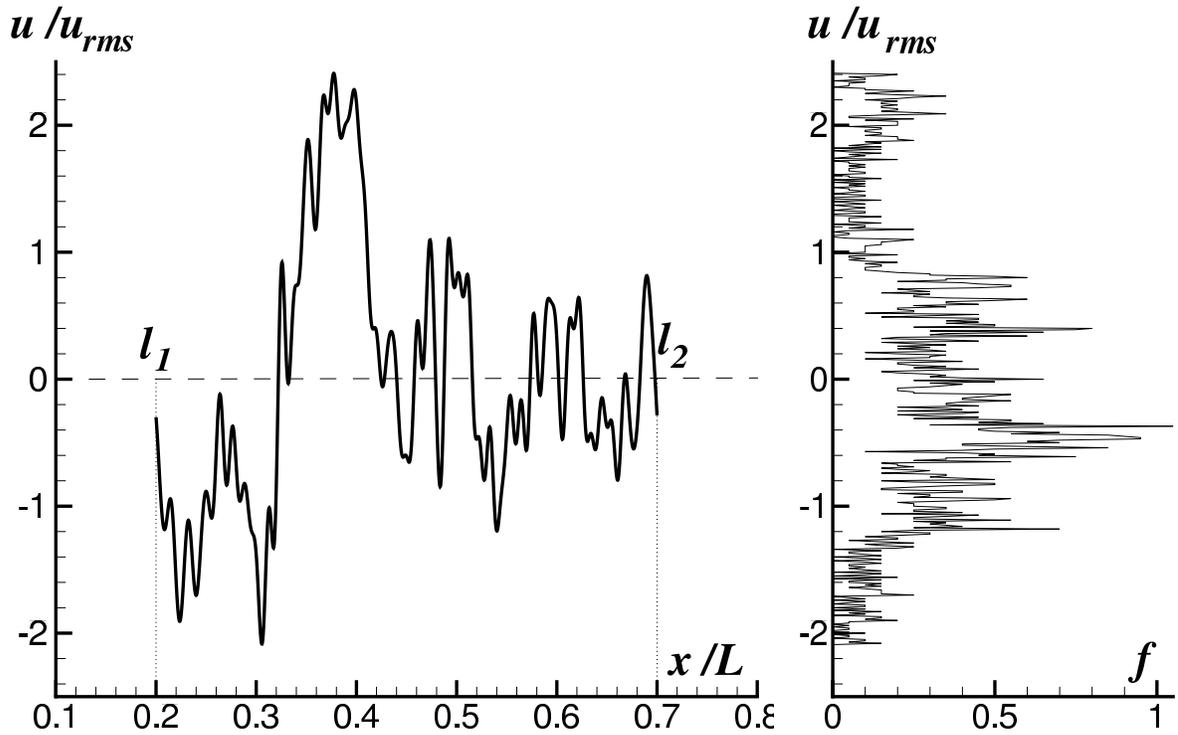}\hfill\mbox{}
\caption{
A scheme for recalculating the instantaneous velocity field
$u(x)$ {\bf (a)}
to the velocity distribution $f_H(u)$ {\bf (b)}.
The recalculation is performed by ``projecting''
the multi-valued function $x(u)$
(the inverse of the function $u(x)$)
onto the $u$ axis.
The $f_H(u)$ profile is similar to the profile
of a hydrodynamically
(without thermal broadening)
broadened line.
\label{fig:fig11}
}
\end{figure}

Let us explain the meaning of the distribution $f_H(u)$.
There is a function $u(x)$.
Let us divide the $u$ axis into segments $\delta u$
that are small compared to the viscous fluctuation amplitude
$(\Delta u)_{\nu}$.
Consider an arbitrary point $u_{\star}$ on the $u$ axis.
By definition,
$$
f_H (u_{\star}) =
d p / d u =
\sum \delta x / l_{21} \, \delta u, \eqno (5.1)
$$
where the differential $dp$ characterizes the ``weight''
of the set of subregions
that move with velocity $u_{\star}$.
We assume that the argument of the function $u(x)$
belongs to the region $l_1 \le x \le l_2$ inside the cluster.
The condition
$
u_{\star} - \delta u/2 \le
 u(x_{\star}) \le
u_{\star} + \delta u/2
$
is satisfied in each of the small segments $\delta x$ in sum (5.1).
In this condition,
the points $x_{\star}$
belong to any of the small segments $dx$ in sum (5.1).
Because of the statistical pulsations in velocity $u(x)$,
sum (5.1) can contain many terms.
Their number is approximately equal
to the numbers of intersections of the function $u = u(x)$
with the straight line $u = u_{\star}$.
It is easy to see that distribution (5.1) is normalized,
$\int_{-\infty}^{\infty} f_H \, du = 1$.
The corresponding illustration is presented in Fig. \ref{fig:fig11}.

Let us first consider an example
with a monotonic function $u(x)$
and then return to the non-monotonicity
and multivaluedness of the function $x(u)$
that is the inverse of $u(x)$
and to the turning points.
In the monotonic case,
$$
1 / [\, d u(x)/dx \,] = d x(u) / du = l_{21} f_H(u);
$$
i.e., the velocity gradient determines $f_H$.
The equations $u = u(x)$
and $f_H = \left[ l_{21} \frac{ du(x)}{ dx } \right]^{-1}$
parametrically specify the distribution $f_H(u)$.
Eliminating the parameter $x$ from this pair of equations
yields an explicit dependence $f_H(u)$.

Let $u(x) = U \sin(2 \pi x/L)$.
Eliminating the parameter $x,$ we then obtain
$$
f_H (u) =
\left(\, l_{21} \, du / dx \,\right)^{-1} =
\frac{L}{2\pi l_{21}} \,
\frac{ 1 }{U \cos(2\pi x/L)} =
\frac{L}{2\pi l_{21}} \,
\frac{ 1 }{ \sqrt{ U^2 - u^2 }}. \eqno (5.2)
$$
Expression (5.2) specifies the dependence $f_H (u)$
in explicit form.
Of course,
an explicit expression
can be derived only for very simple functions $u(x)$.

The example with an oscillating function $u(x)$
clearly illustrates an interesting singularity
of the distribution $f_H$ in the cold\footnote{
In Section 8 of the full paper,
we show how thermal Doppler broadening ``smears'' the singularities
caused by turning points.}
case.
It becomes infinite\footnote{
Of course,
this is an integrable singularity.}
at the stop, turning, or cuspidal points
of the inverse function $x(u)$
that are the extrema of the function $u(x)$;
because of the root behavior,
the distribution $f_H(u)$ is asymmetric
about the singularity.
The case with one harmonic (5.2)
is an example that greatly differs from the Gaussian distribution.
In this example,
the dependence $f_H(u)$ has a minimum
at the center and a singular break in the distribution tail.
Consequently, in this example,
the regions
in which there is motion relative to the observer
are more representative
than the regions at rest
(that is why the distribution maximum is off the center).
As we will see below,
in the case of many harmonics,
the turning points or extrema
lead to a needle-shaped distribution $f_H(u)$.

Thus, it is clear that distribution (5.1) is,
in a sense,
obtained by ``projecting'' the inverse function $x(u)$
onto the $u$ axis,
as shown in Fig. \ref{fig:fig11}
(the connection between Figs. \ref{fig:fig11} (a) and (b)).
Fig. \ref{fig:fig11} (a) shows a realization of $u(x)$ (4.1)-(4.3)
with random phases $\psi_n\,$
$(N \le n \le N_{\nu},\,$
$N_{\rm mix} = 2, N_{\nu} = 100)$
in the interval $l_1 \le x \le l_2\,$
$(l_1 = 0.2 L, l_2 = 0.7 L)$.
The viscous limit is fairly close:
$
(\Delta x)_{\nu} = l_{21}/N_{\nu} \sim 10^{-2} L,\,$
$(\Delta u)_{\nu} \sim 0.1 u_{\rm rms}$.
The viscous values determine the heights and widths
of the $u(x)$ spikes (Fig. \ref{fig:fig11}).

In calculating the distribution $f_H(u)$,
we substituted the velocity $u(x)$
with the set of its values $U_i$ at points $x_i = i\delta x$
(discretization).
The $\delta x$ step was chosen to be small
compared to the viscous length $(\Delta x)_{\nu}$
in order to resolve dissipating eddies.
The distribution $f_H(u)$ was calculated numerically.
It was replaced by a histogram (Fig. \ref{fig:fig11} (b))
with a small velocity step $\delta u \ll (\Delta u)_{\nu}$
in order not to lose any fluctuation.
Because of the discretization,
the above singularities of the distribution $f_H(u)$
are cut off at turning points.
The height of the spikes in the distribution $f_H(u)$
is on the order of $\sqrt{(\Delta u)_{\nu}/\delta u},$;
cf. (5.2).
The value of $f(u_{\star})$ at the step
$$
u_{\star}-\delta u/2 \leq u \leq u_{\star}+\delta u/2
$$
of the histogram
is proportional to the number of discretization points $U_i$
at the limits $u_{\star}\pm\delta u/2$.
After the normalization $\int f_H \, du=1$,
we obtain the distribution $f_H$ shown in Fig. \ref{fig:fig11} (b).

It would be natural
to analyze the deviations from the Gaussian distribution
together with the calculation of moments,
because the extent
to which the distribution is non-Gaussian
can be characterized by the deviations of moments
from their Gaussian values.

\section{The Method of Moments}

Above,
we described the procedure for an exact calculation
of the distribution $f_H$
(the example in Fig. \ref{fig:fig11}).
Concurrently,
it is instructive to describe the method for its analytical
calculation.
We approximate the distribution  $f_H$ by a polynomial
$$  
f(u)=a_0 +a_1 u +a_2 u^2 + ... =
\sum_{j=0}^{n_{app}} a_j u^j. 
\eqno (6.1)
$$
The calculation procedure
is based on the fact that the moments
are determined by two independent methods.
Let us calculate the moments $\mu_n$
directly from the velocity field $u(x)$
and the distribution $f_H (u)$.
By definition,
$$
\mu_n =
\int_{l_1}^{l_2} [u(x)]^n \, dx/l_{21}
=\int_{-\infty}^{\infty} u^n \,
f(u) \, du.
\eqno (6.2)
$$
The first three moments are known
(do not depend on the phases $\psi_n$):
$\mu_0 = 1$
(the normalization condition $\int f_H \, du = 1$),
$\mu_1 = 0$ (centering (4.2)),
and $\mu_2 = 1$
(the choice of velocity unit (4.3)).

We truncate series (6.1)
at the approximation order $n_{\rm app}$.
Expansion (6.1) contains $n_{\rm app}$ + 1 unknown coefficients
$a_n$.
Let us set up a system of equations to determine them.
The equations are $\mu_n^u = \mu_n^f$,
where the expressions for $\mu_n^u$ and $\mu_n^f$
are calculated from $u$ and $f_H$, respectively (see (6.2)).
Let us
first calculate the missing moments
$\mu_3^u, \mu_4^u,$, ..., $\mu_{n-app}^u$
from the velocity $u(x)$.
For a given realization,
these are just some specific numbers.
We then express integrals (6.2),
$$
\mu_n^f = \int_{u_l}^{u_r} u^n \, f(u) \, du,
\eqno (6.3)
$$
in terms of expansion (6.1).

Moments (6.3) are linear forms in unknowns $a_n$.
The coefficients of the forms
depend on the left, $u_l < 0$,
and right, $u_r > 0$,
integration limits in (6.3).
Clearly, the power-law approximation (6.1)
does not describe the decaying tails of the distribution $f_H$.
Polynomial (6.1) intersects the $u$ axis at points $u_l$ and $u_r$.
Inside the segment $u_l \le u \le u_r$,
the polynomial is positive.

Solving the linear system of equations
$\mu_n^u = \mu_n^f$
for the unknowns $a_n$,
we express $a_n$ in terms of the zeros $u_l$ and $u_r$
of the polynomial $f_H$ in the form of a rational\footnote{
The ratio of polynomials}
function of $u_l$ and $u_r$.
Substituting the derived expressions for $a_n (u_l, u_r)$
into (6.1) yields the distribution $f_H (u_l,u_r;u)$.
The zeros $u_l$ and $u_r$
are defined by a system of two equations
$
f_H(u_l, u_r; u_l)$ = 0
and
$
f_H(u_l, u_r; u_r)$ = 0.
At $n_{\rm app}$ = 2
(approximation (6.1) is a parabola),
$u_l = - \sqrt{5}$ and $u_r = \sqrt{5}$.
In high orders, the equations rapidly become cumbersome.

\begin{figure}
\mbox{}\hfill\epsfxsize 1\hsize\epsfbox{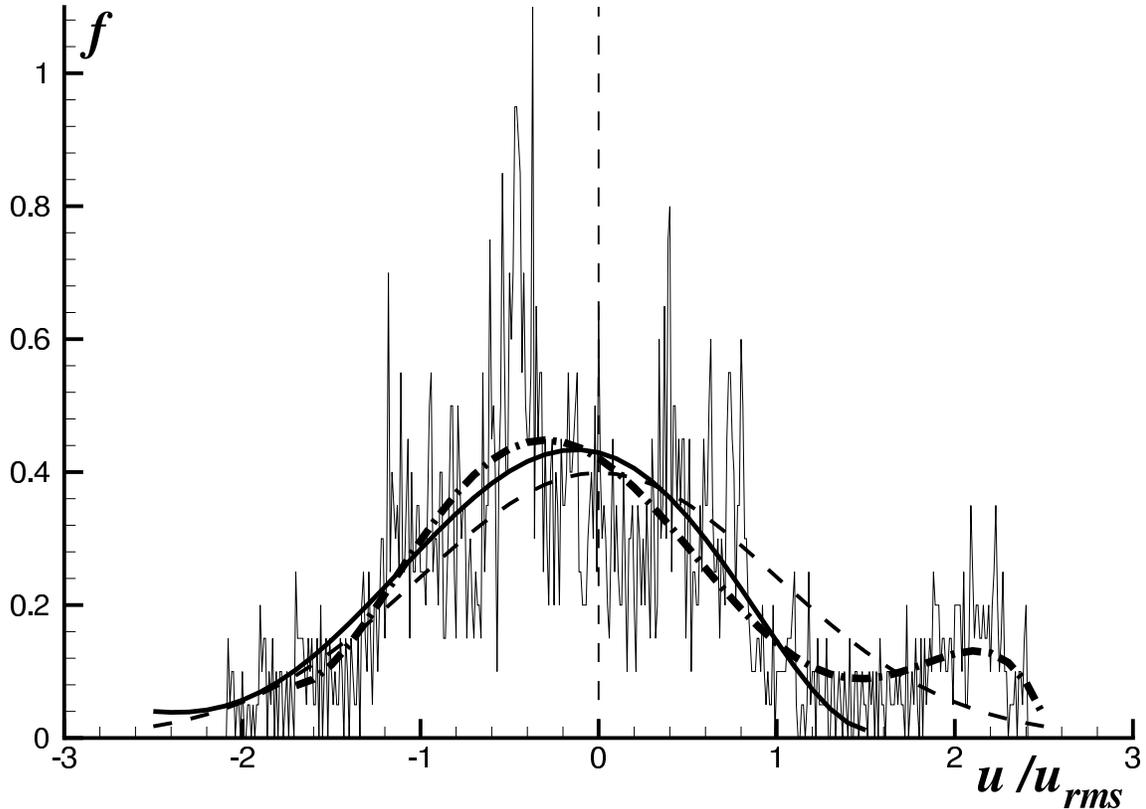}\hfill\mbox{}
\caption{
Comparison of the distributions $f_H(u):$
the thin fluctuating solid curve -- exact calculation,
the thick smooth solid curve -- calculation by the methods of moments
(a polynomial up to $u^5$ inclusive),
the thick dot-dashed curve -- a polynomial up to $u^6,$
the thin dashed curve -- Gaussian distribution $f^G.$
The distribution $f^G$ is symmetric about the vertical line $u=0.$
\label{fig:fig12}
}
\end{figure}

Examples of polynomials for $n_{\rm app}$ = 5 and 6
are given in Fig. \ref{fig:fig12}.
They refer to the realization of $u(x)$
shown in Fig. \ref{fig:fig11}.
The polynomials satisfactorily fit the exact distribution $f_H$.
The sixth-order polynomial
(heavy dash-dotted line)
catches even the additional peak in the right wing.
Of course, polynomials of a limited order
smooth out the ``needles'' of the exact distribution.
The Gaussian distribution is indicated in Fig. \ref{fig:fig12} by the dashed curve.
We see the essentially non-Gaussian behavior,
which manifest itself in the profile asymmetry
relative to the dashed $u = 0$ vertical line (Fig. \ref{fig:fig12})
and in the break of the distribution tails
and the appearance of additional peaks.
In addition, the odd moments are nonzero,
and the ratios of sequential even moments
differ markedly from the Gaussian ratios.
For the typical realization shown in Fig. \ref{fig:fig11},
$\mu_3$ = 0.51
and $\mu_5$ = 2.97.
The measure of smallness of the odd moments
is determined by the normalization $\mu_2$ = 1 (4.3).
The increase in the absolute value of the moment
with increasing moment number
should also be taken into account.
In the Gaussian case,
the following relations hold:
$
\mu_4^G = 3 \, (\mu_2^G)^2,
$
$
\mu_6^G = 15 \, (\mu_2^G)^3
$.
In our case,
however, $\mu_4/3 = 0.93$ and  $\mu_6/15 = 0.71$.

Let us consider the influence of viscosity.
To assess the role of the width of the inertial range
log$_{10} \frac{N_\nu}{N_{\rm mix}}$,
we compared the velocities $u(x)$ and the distributions $f_H(u)$
with $N_\nu \sim 10^2$
and  $N_\nu \sim 10^3$.
As should be the case,
the differences are attributable
to the appearance of small-scale fluctuations
as $N_\nu$ (3.5) greatly increases.
The positions and shapes of the significant fluctuations
with $\Delta_u \sim 0.1 u_{\rm rms}$
depend weakly on $N_\nu$
if $N_\nu > 100$.
The first moments
(up to the sixth moment inclusive)
change only slightly (by $\sim$ 1\%).

\section{Gaussian Asymptotics}

The main regulator
on which the deviation of $f_H$
from the Gaussian distribution $f^G$ depends
is the parameter $N_{\rm mix}$
that specifies the extent
to which the scale of turbulence (3.1) is large.
The viscous scale in this ratio is unimportant.
Therefore,
we set $N_\nu = \infty$.
The difference between the distributions $f_H$
and $f^G$
decreases with increasing $N_{\rm mix}$.
At $N_{\rm mix} = \infty$, the profile is Gaussian.
Let us first show this
and then present the results
that illustrate the pattern of convergence
$f_H \rightarrow f^G$ with increasing $N_{\rm mix}$.

Let us analyze the case with $N_{\rm mix} \gg 1$.
Above,
the calculations were associated with the dependence of velocity $u$
on coordinate $x$.
In this case, the phases $\psi_n$ (4.1)
were fixed.
This is an approach
with averaging over the coordinate.
Let us look at the question differently.
Let $x$ be fixed
($x = x_{\rm fix}$).
The quantity
$
u(x_{\rm fix}; \psi) = u(\psi)
$
gives the velocity at point $x_{\rm fix}$
for a given set of phases $\psi$ (4.1).
Let us study the statistics of $u(\psi)$
(an approach with phase averaging).

Since the phases $\psi_n$ are independent and random
and the position $x$ is fixed,
we may omit the regular oscillating factors
exp ($i 2\pi n x/L$) in (4.1).
Consequently,
the velocity
$
u(x_{\rm fix};\psi)
$
is given by
$$
u =
(\, N_{\rm mix} \,)^{1/3}\sum_{n = N_{\rm mix} }^{\infty}
\xi_n/n^{5/6},
\eqno (7.1)
$$
where,
in place of the phases $\psi_n$
we introduced the equivalent independent random variables
$\xi_n\,$
$
\langle \xi_n \rangle = 0,
$
and
$
\langle \xi_n^2 \rangle = 1$;
$\, \langle \rangle$ denotes averaging.
We assume, for simplicity,
that $\psi_n$  takes on values of +1 or -1
with equal probabilities.
The probability distribution function $f_H(u)$
of the random variable $u$ (7.1) should be determined.

We see from definition (7.1)
that the mean velocity
$\langle u \rangle$
is equal to zero.
The probability distribution $f_H(u)$
is symmetric
($f_H$ is an even function of $u$).
All odd moments are zero\footnote{
In the specific realizations
considered in Sections 4-6,
the function  $f_H(u)$ is asymmetric.
Here,
we deal with the probability distribution
after averaging over all possible phases
(i.e. over all possible realizations).
This distribution is symmetric.}.
It is easy to show that the second moment of (7.1) is
$$
\langle u^2 \rangle =
(\, N_{\rm mix} \,)^{2/3}
\sum_{ n = N_{\rm mix} }^{\infty}
n^{-5/3} =
(\, N_{\rm mix} \,)^{2/3} \, \zeta(5/3, N_{\rm mix}),
\eqno (7.2)
$$
where $\zeta$(5/3, $N_{\rm mix}$)
is the generalized Riemann zeta function (Vinogradov 1979).
In the limit $N_{\rm mix} \gg 1$,
$\, \langle u^2 \rangle = 3/2$,
because
$\sum n^{-5/3} \approx \int dn/n^{5/3}$.

Let us calculate the fourth moment.
We have
$$
u^4/(\, N_{\rm mix} \,)^{4/3} =
( \, \sum \xi_n/n^{5/6} \,)^4 =
  - 2 \sum \xi_n^4/n^{10/3} +
 3 \, ( \, \sum \xi_n^2/n^{5/3} \,)^2 + A.
\eqno (7.3)
$$
In (7.3),
the term $A$
is the sum of the terms
with an odd number of factors $\xi_n$.
After averaging, the term $A$ vanishes,
and we obtain
$$
\langle u^4 \rangle/(\, N_{\rm mix} \,)^{4/3} =
 - 2 \sum n^{-10/3} +
   3 \, ( \, \sum n^{-5/3} \,)^2 =
 - 2 \, \zeta(10/3, N_{\rm mix}) +
   3 \, [\, \zeta(5/3, N_{\rm mix}) \,]^2,
\eqno (7.4)
$$
because $\xi_n = \pm 1$
and $\xi_n^4 = 1$.
For $N_{\rm mix} \gg 1$,
expression (7.4) tends to the limit
$$
\langle u^4 \rangle = 3 \, (3/2)^2.
\eqno (7.5)
$$
As we see,
the ratio of the fourth moment (7.5)
to the square of the second moment (7.2),
$\langle u^4 \rangle /(\, \langle u^2 \rangle \,)^2$,
is equal to three.
This should be the case for the Gaussian distribution.
The aforesaid is also true for the random variables $\xi_n$
of a more general form
than $\xi_n = \pm 1$.
Indeed,
for $N_{\rm mix} \gg 1$,
sum (7.1) includes many approximately equal
(in absolute value)
independent random terms.
In this case,
in view of the central limit theorem,
the probability distribution of the random variable $u$
is Gaussian.
At $N_{\rm mix} \sim 1$,
this is not the case,
because th first terms in sum (7.1)
significantly differ
due to the factor $n^{-5/6}$.

\vspace{0.4cm}

\centerline{\bf Table 2.}

\begin{center} \begin{tabular}{|c|c|c|c|c|c|c|} \hline 
$N_{mix}   $&$\mu_2-1$&$\mu_3$&$R_4-1
$&$R_4^{\zeta}-1$&$\mu_5$&$R_6-1$\\ \hline
$2\;(\star)$&$  0    $&$0.51 $&$-0.07 $&$             $&$2.97
$&$-0.29$\\ \hline
$2         $&$  0    $&$0.31 $&$-0.11 $&$-0.078       $&$4.8
$&$-0.21$\\ \hline
$ 10       $&$  0    $&$0.21 $&$-0.077$&$-0.013       $&$1.7
$&$-0.23$\\ \hline
$ 50       $&$  0    $&$0.014$&$-0.045$&$-0.0026      $&$0.25
$&$-0.16$\\ \hline
$\infty    $&$  0    $&$0    $&$0     $&$0            $&$0    $&$ 0
$\\ \hline
\end{tabular}\end{center}

It remains to study the convergence $f_H \rightarrow F^G$
with increasing $N_{\rm mix}$.
The corresponding results are presented in Table 2.

This table shows
how the moments converge to their Gaussian values
(the last row).
The moments $\mu_n$ are defined by formulas (6.2),
$
R_4 = \mu_4/3\mu^2_2,
$
$\, R_6 = \mu_6/15 \mu_2^3$.
The first moment is equal to zero due to centering (4.2).
Because of normalization (4.3),
the second moment is always equal to unity.
In the Gaussian limit,
$\mu_3 = \mu_5 = 0$ and $R_4 = R_6 = 1$.
The procedure for calculating the moments $\mu_3 - \mu_6$
was described in Section 5 and 6.
The table list the mean values
obtained by averaging over many realizations $\vec \xi, u(x)$ (4.1).
For the odd moments,
we average the absolute value of the moment
to establish the degree of deviation from zero.
The realization-averaged values
are given in rows 3-5 of Table 2.
For comparison with a unit realization,
the second row gives the values that refer to the example
 in Fig. \ref{fig:fig11};
see also the end of Section 6.
This row is highlighted by the symbol ($\star$).
The formula
\[
R_4^{\zeta}( N_{\rm mix} ) - 1 =
\langle u^4 \rangle/ 3 ( \, \langle u^2 \rangle \, )^2 - 1 =
(-2/3) \,\zeta( 10/3, N_{\rm mix} )/[\,\zeta(5/3, N_{\rm mix})\,]^2,
\]
used in column 5 of the table
follows from expressions (7.2) and (7.4).
It shows the deviation from the Gaussian distribution
in the case of averaging over all realization $\xi_n$.
The corresponding $R_4$ calculated form the zeta function
is denoted by the superscript $\zeta$.

Let us discuss the data in the Table 2.
We see that the moments approach their Gaussian values
as $N_{\rm mix}$ increases
(refinement of the leading or dominant turbulence scale (3.1)).
Therefor, the distribution $f_H$
also tends to $f^G$.
As should be the case,
the convergence in lower moments ($\mu_3, \mu_4$)
is faster
than the convergence in higher moments ($\mu_5, \mu_6$).
The deviations in higher moments are larger,
and they decrease with increasing $N_{\rm mix}$ more slowly.
The main conclusion
is that for large-scale turbulence
with $N_{\rm mix} = 2-10$
$\, (L_{\rm mix} \sim L$, (3.1)),
there are significant deviations from the Gaussian distribution.

\section*{References}


\noindent E. Churazov, M. Br\"uggen, C.R. Kaiser, et al.
Astrophys. J. {\bf 554}, 262 (2002a).

\noindent E. Churazov, R. Sunyaev, W. Forman, and H. Boehringer,
Mon. Not. R. Astron. Soc. {\bf 332}, 729 (2002b).

\noindent E. Churazov, W. Forman, C. Jones, R. Sunyaev, and H. Boehringer,
Mon. Not. R. Astron. Soc. (2003 accepted);
astro-ph/0309427.

\noindent A.C. Fabian, J.S. Sanders, C.S. Crawford, C.J. Conselice,
J.S. Gallagher, and R.F.G. Wyse,
Mon. Not. R. Astron. Soc.
{\bf 344,} L48 (2003).

\noindent C.S. Frenk, S.D.M. White, P. Bode, et al. Astronphys. J. {\bf
525}, 554 (1999).

\noindent J.P. Ge and F.N. Owen, Astron. J. {\bf 105}, 778 (1993).

\noindent M. R. Gilfanov, R.A. Sunyaev and E.M. Churazov,
Sov. Astron. Lett {\bf 13}, 3 (1987).

\noindent V.V. Ivanov, {\it Radiative Transfer and the Spectra of
Celestial Bodies}, Nat. Bureau of Standards Spec. Publ. No. 385 (1973).

\noindent R.K. Janev, L.P. Presnyakov, and V.P. Shevelko, {\it Physics of
Highly Charged Ions. Springer Series in Electrophysics} (Berlin:
Springer, 1985), Vol. 13.

\noindent A.N. Kolmogorov, Dokl. Akad. Nauk SSSR 30, 299 (1941).

\noindent L.D. Landau and E.M. Lifshitz, {\it Hydrodynamics} (Nauka,
Moscow, 1986) [in Russian].

\noindent {\it Mathematical Encyclopaedia,} Ed. by I.M. Vinogradov
(Sov. Encyclopedia, Moscow, 1979), Vol. 2 [in Russian].

\noindent A. S. Monin and A.M. Yaglom {\it Statistical Hydromechanics} (
Nauka, Moscow, 1965) [in Russian].

\noindent D. Nagai, A. Kravtsov, and A. Kosowsky, Astrophys. J. {\bf 587},
524 (2003)/

\noindent M.L. Norman and G.L. Bryan, {\it The Radio Galaxy Messier} 87,
Ed. by H.-J. Roeser and K. Meisenheimer (Springer, Berlin, 1999),
Lecture Not. Phys. {\bf 530}, ISSN0075-8450.

\noindent F.S. Porter and K. Mitsuda (Astro-E2/XRS Collab.) {\it American
Astron. Soc. HEAD Meeting} No. 35, No. 33.05 (2003).

\noindent M.N. Rosenbluth and R.Z. Sagdeev, {\it Handbook  of Plasma
Physics} (North-Holland, Amsterdam, 1983).

\noindent R.K. Smith, N.S. Brickhouse, D.A. Liedahl, and J.C. Raymond,
Astrophys. J. {\bf 556} L91(2001).

\noindent R.A. Sunyaev, Astron. Lett. {\bf 3} 268 (1977).

\noindent R.A. Sunyaev, M. Norman, and G. Brian,
Astronomy Letters {\bf 29} 783 (2003);
astro-ph/0310041.

\noindent R.A. Sunyaev and Ya.B. Zeldovich,
Astrophys. Space Sci. {\bf 7}, 3 (1970).

\noindent R.A. Sunyaev and Ya.B. Zeldovich,
Mon. Not. R. Astron. Soc. {\bf 190}, 413 (1980).


\end{document}